\documentclass[journal,hidelinks]{new-aiaa}
\usepackage[utf8]{inputenc}
\usepackage{textcomp}

\usepackage{graphicx}
\usepackage{amsmath}
\usepackage[version=4]{mhchem}
\usepackage{siunitx}
\usepackage{longtable,tabularx}
\usepackage{comment}
\usepackage[dvipsnames]{xcolor}
\usepackage{lineno}

\title{Constrained re-calibration of Reynolds-averaged Navier-Stokes models}

\author{Yuanwei Bin\footnote{Visiting scholar, Ph.D. Candidate, Mechanical Engineering}}
\affil{Pennsylvania State University, State College, Pennsylvania, USA, 16802}
\affil{State Key Laboratory for Turbulence and Complex Systems, Peking University, Beijing, China, 100871}
\author{George Huang\footnote{{Professor}, Mechanical and Materials Engineering}}
\affil{Wright State University, Dayton, Ohio, USA, 45435}
\author{Robert Kunz\footnote{ Professor, Mechanical Engineering}, Xiang I A Yang\footnote{Assistant Professor, Mechanical Engineering}}
\affil{Pennsylvania State University, State College, Pennsylvania, USA, 16802}

\begin{document}

\maketitle


\begin{abstract}
The constants and functions in Reynolds-averaged Navier Stokes (RANS) turbulence models are coupled.
Consequently, modifications of a RANS model often negatively impact its basic calibrations, which is why machine-learned augmentations are often detrimental outside the training dataset.
A solution to this is to identify the degrees of freedom that do not affect the basic calibrations and only modify these identified degrees of freedom when {\color{black}re-calibrating} the baseline model to accommodate a specific application.
{\color{black}This approach is colloquially known as the ``rubber-band'' approach, which we formally call ``constrained model re-calibration'' in this article.}
To illustrate the efficacy of the approach, we identify the degrees of freedom in the Spalart-Allmaras (SA) model that do not affect the log law calibration. 
By subsequently interfacing data-based methods with these degrees of freedom, we train models to solve historically challenging flow scenarios, including the round-jet/plane-jet anomaly, airfoil stall, secondary flow separation, and recovery after separation.
In addition to good performance inside the training dataset, the trained models yield similar performance as the baseline model outside the training dataset.
\end{abstract}

\section*{Nomenclature}

\noindent(Nomenclature entries should have the units identified)

{\renewcommand\arraystretch{1.0}
\noindent\begin{longtable*}{@{}l @{\quad=\quad} l@{}}
$A$ & model variable\\
$B$ & model variable\\
$c$   & chord \\
$c_{b1}$, $c_{s1}$, $c_{s2}$ & model tunable constant\\
$C_p$& pressure coefficient \\
$C_f$& skin friction coefficient \\
$C_L$& lift coefficient \\
$c_{b2}$, $c_w$ & model fixed constant\\
$d$ & distance to the wall\\
$D$ & diffusion term\\
$D_j$ & diameter for the round jet\\
${\rm D}/{\rm D}t$ & material derivative\\
$f_{\nu1}$, $f_{\nu2}$, $f_w$ & model function\\
$F_{\rm lift}$ & lift\\
$h$ & height of the back-facing step\\
$k$ & turbulent kinetic energy\\
$l_{1/2}$ & distance from the jet centerline to half-velocity location\\
$L$ & length of the prolate spheroid\\
$P$ & pressure\\
$r$ & model variable\\
$Re_x$ & Reynolds number based on $x$\\
$Re_\tau$ & Reynolds number based on $u_\tau$\\
$Re_c$ & Reynolds number based on $c$\\
$Re_L$ & Reynolds number based on $L$\\
$S$ & spreading rate\\
$t$ & time\\
$u$, $U$ & velocity\\
$U_{\rm inf}$ & freestream velocity\\
$x_i$ & $i$-th Cartesian direction\\
$y$ & vertical location to the wall\\
$\alpha$ & angle of attack\\
$\beta$, $\beta^*$ & model constant\\
$\gamma$ & model constant\\
$\epsilon$ & dissipation rate\\
$\kappa$ & von K{\'a}rm{\'a}n constant\\
$\mu$ & fluid dynamic viscosity\\
$\mu_t$ & turbulent eddy dynamic viscosity\\
$\nu$ & fluid kinematic viscosity\\
$\nu_t$, $\tilde{\nu}$ & turbulent eddy viscosity\\
$\rho$ & fluid density\\
$\sigma$, $\sigma^*$ & model constant\\
$\tau$ & shear stress\\
$\chi$ & model variable\\
$\omega$ & specific frequency\\
\end{longtable*}}

\section{Introduction}\label{sec:intro}

Reynolds-averaged Navier Stokes (RANS) is an extensively-used tool in fluid engineering \cite{durbin2018some}.
It solves for the mean flow and models the entirety of turbulence \cite{pope2000turbulent}.
Due to the high cost of scale-resolving simulation at high Reynolds numbers and the limited computational resources \cite{chapman1979computational,yang2021grid}, non-scale resolving tools like RANS will continue to be the workhorse for fluid engineering \cite{li2022grid,chapman1979computational}.
It is, therefore, justified to study new approaches to RANS modeling, {\color{black}e.g., data-based approaches.}

A RANS model represents a mapping between the mean flow and the underlying turbulence.
However, due to the stochastic nature of turbulence, manifested as multiple states \cite{yang2021bifurcation,ravelet2004multistability,xie2018flow,weiss2010finite} and non-ergodicity, there is not a general mapping between the mean flow and turbulence, and therefore no general RANS model \cite{spalart2015philosophies}.
Consequently, RANS modeling has an element of art, giving rise to a large number of RANS models with varying ranges of applicability \cite{spalart1992one,menter1994two,wilcox1988reassessment,wilcox1998turbulence,wilcox2008formulation,chien1982predictions,launder1983numerical,abdol2016verification,durbin1991near}.
Despite the proliferation of RANS models, model re-calibration is a recurring theme.
The process of model re-calibration often involves adjusting constants, functions, and terms according to some existing reference data.
This process is conventionally ad-hoc, relying heavily on users' experiences \cite{huang2023distilling,zhang2020methodology, parente2011improved}.
Examples include the calibration of the $k$-$L$ model for Richtmyer-Meshkov mixing instability \cite{zhang2020methodology}, the adjustment of the $k$-$\epsilon$ model for atmospheric flows \cite{cindori2018steady, parente2011improved}, and the re-calibration of the RNG model, the SA model, and the $k$-$\epsilon$ model for urban canopies \cite{gimenez2019optimization}.
More recently, machine learning tools were used to help re-calibrate RANS models.
For example, the methods in Ref. \cite{ling2016machine,ling2016reynolds, wang2017physics,wu2018physics} help re-calibrate the Reynolds stress;
the methods in Ref. \cite{singh2016using,singh2017machine} help re-calibrate the production term;
the methods in Ref. \cite{fang2023toward,zhao2020rans} help re-calibrate the terms in the Reynolds stress.
{\color{black}
By applying these methods, one recalibrates the terms in an existing model.
Take the field inversion and machine learning (FIML) method as an illustrative example \cite{duraisamy2019turbulence,singh2016using,singh2017machine}.
The method re-calibrates a target term, usually the production term, in the auxiliary transport equation in a RANS model through a prefactor.
This re-calibration process gives rise to pre-factors that are far from 1 \cite{rumsey2022search}.
While the FIML has yielded good results for turbomachinery \cite{ferrero2020field}, Ahmed body \cite{wu2023enhancing}, among others, the method, like all other ML methods, does not generalize.
Furthermore, the re-calibration often negatively impacts the basic calibrations of a baseline model. 
This is undesirable.
Irrespective of how well a model does for, e.g., airplanes, it should preserve basic calibrations like the law of the wall \cite{rumsey2022search, spalart2015philosophies}.
}

{\color{black}
This deterioration of generalizability is not hard to fathom. 
The terms and the model constants in a RANS model are coupled, and therefore adjusting one term or one constant affects, often negatively, all calibrations of the model.
Although certain calibrations are not critical, basic calibrations like the law of the wall are regarded as critical to a model's ability to generalize \cite{spalart2015philosophies,spalart2023old}.
The goal of the present work is to provide constraints that help preserve these basic RANS model calibrations in the process of machine learning.
The method we propose here is ``constrained re-calibration''.
The approach is colloquially known as the ``rubber band (RB)'' approach (among attendees of the Turbulence Modeling Resources meeting series).
}
A RB RANS model would contain adjustable constants.
These adjustable constants control the model's behaviors in complex flows but do not affect designated basic calibrations.
By adjusting these constants, users would be able to calibrate the baseline model to accommodate specific needs while preserving the baseline model's generalizability.

In the present literature, the Generalized $k$-$\omega$ (GEKO) model seems to be the only RB model  \cite{menter2019best}.
The GEKO model was developed in 2019.
{\color{black}Menter's goal then was to consolidate the many two-equations models: the two-equation $k$-$\epsilon$ model \cite{launder1983numerical} and $k$-$\omega$ model \cite{menter1994two,wilcox1988reassessment,wilcox1998turbulence,wilcox2008formulation}, which are not fundamentally or conceptually different.}
GEKO has six adjustable constants that can be freely varied without compromising the logarithmic law.
The model has received much attention since its release \cite{strokach2021simulation,szudarek2022cfd,jung2021uncertainty,sharkey2019numerical,nair2022resistance}:
Strokach et al. and Szudarek et al. employed GEKO and computed flow in rocket and train applications \cite{strokach2021simulation, szudarek2022cfd};
Jung et al. \cite{jung2021uncertainty} utilized Bayesian inference and calibrated the model constants for high-speed flows. 
However, the details of the model are not open to the public and are kept proprietary to ANSYS, the company that developed it. 
Consequently, its use is limited to the ANSYS software, and there have been limited developments following this line of thought.
Additionally, the question remains open as to whether the RB approach applies to two-equation models only.

This work will present a modeling framework that allows constrained re-calibration of the SA model.
The reformulated model has two groups of adjustable constants, with two adjustable constants in each group.
The two constants in the first group control the model's behaviors in unbounded flows, and the two constants in the second group control the model's behaviors in separated flows.
We will re-calibrate the SA model to accommodate flow scenarios that are historically challenging for RANS models, including plane-jet/round-jet anomaly, secondary flow separation, stall, and flow recovery (after separation) \cite{bridges2010establishing,bridges2011nasa,barri2010dns,seifert2002active,xiao2007prediction}. 
More importantly, we will show that these ``re-calibrations'' are robust and have non-detrimental behaviors outside the training dataset.

Before we proceed further, we make the following distinctions.
The first distinction is between ``a model'' and ``a modeling approach''.
A model is specific.
A user can pick up a model as is and use it for predictive modeling.
A modeling approach is less specific.
It refers to the framework or methodology used to create models.
For example, FIML in Ref. \cite{duraisamy2019turbulence,singh2016using} is a modeling approach.
It is a framework that allows one to construct an augmentation to a baseline model from data.
In addition to FIML, Tensor-Based Neural Networks (TNBB) in Ref. \cite{ling2016machine}, Phyiscs-Informed Machine Learning (PIML) in Ref. \cite{wang2017physics,wu2018physics}, among others \cite{zhao2020rans,fang2023toward}, are all modeling approaches.
Constrained recalibration, the focus of this paper, is also a modeling approach. 
Although one can only show the effectiveness of a modeling approach by studying specific models, the focus of the above-cited studies and the present work is not on a specific model.

Next, we distinguish between ``where'' and ``how'' a model re-calibration happens.
To illustrate these two concepts, we take, again, FIML as an example.
FIML's augmentation is an augmentation to the production term---this is {\it where} the re-calibration happens.
The re-calibration process itself involves field inversion and machine learning, which are {\it how} the re-calibration happens.
Most existing work on the topic of machine-learned RANS models focuses on {\it how} rather than {\it where}:
the focus of Ref. \cite{ling2016machine} is on preserving Galilean invariance in a neural network, which is {\it how}, and
the focus of Ref. \cite{wu2018physics} is on incorporating all information as inputs, which is also {\it how}.
Minimal attention has been given to {\it where}.
Instead of focusing on the {\it how} aspect, the present paper focuses on the {\it where} aspect.
{\color{black}By identifying degrees of freedom that do not affect basic calibrations and addressing the {\it where} aspect, we will see that even simple methods, e.g., Bayesian optimization, can lead to improvements that generalize outside the training set.}

{\color{black}Lastly, we distinguish between the rubber-band approach, the zonal approach, and constrained re-calibration. In this article, we use constrained re-calibration and constrained model re-calibration, as well as rubber band approach, interchangeably. The purpose of constrained re-calibration, as we have discussed, is to preserve the basic calibrations of a RANS model. The method, as we will elaborate, involves identifying degrees of freedom that do not affect basic calibrations. Model calibration then involves only these degrees of freedom. The true strength of this method is evident when a RANS practitioner employs one constrained re-calibration in one region and another in a different region of a complex flow. This is what we refer to as the zonal approach. It's important to note that in this context, the zonal approach is not limited to the treatment of the near-wall region. Although this paper does not concern zonal models, the ultimate application of rubber-band models is within the context of zonal models.
}

The rest of the paper is organized as follows.
In section \ref{sec:method}, we present the details of a reformulated SA model.
In section \ref{sec:results}, we show the results.
Last, we conclude in section \ref{sec:conclusion}.

\section{Methodology} 
\label{sec:method}

In this section, we reformulate the SA model and present the details of the reformulated model in Section \ref{sub:reformulation}.
We will identify degrees of freedom in the reformulated model that do not affect the law of the wall.
This addresses the {\it where} aspect of model recalibration.
We may turn to any existing machine-learning method, e.g., FIML, TBNN, PIML, etc., for the {\it how} aspect.
In this study, we will resort to the simple Bayesian method.
The details of the method are presented in Section \ref{sub:bayesian}.
Last, we discuss the effects of the adjustable constants on the model's behaviors in free-shear and wall-bounded flows in Section \ref{sub:physics}.
There, we will gain some physical understanding of the model.
Physical understandings are often missing in the recent machine learning literature but, we believe, are critical to a user \cite{vadrot2023log,bin2023data}.

\subsection{Reformulation}
\label{sub:reformulation}

Consider the one-equation SA model.
The model is an eddy viscosity one.
It solves the following transport equation for $\tilde{\nu}$:
\begin{equation}
\frac{\rm D \tilde{\nu}}{{\rm D} t}=c_{b1} \tilde{S}\tilde{\nu} -c_{w1}f_w\left(\frac{\tilde{\nu}}{d}\right)^2+\frac{1}{\sigma}\left[\frac{\partial }{\partial x_j}\left((\nu+\tilde{\nu})\frac{\partial \tilde{\nu}}{\partial x_j}\right)+c_{b2}\frac{\partial \tilde{\nu}}{\partial x_j}\frac{\partial \tilde{\nu}}{\partial x_j}\right].
\label{eq:SA}
\end{equation}
{\color{black}
Here, $\tilde{\nu}$ is an auxiliary variable.
It varies linearly with $y$ from the wall through the viscous sublayer to the top of the log layer, i.e., $\tilde{\nu}=u_\tau \kappa y$.
The eddy viscosity is a prescribed function of this auxiliary variable:
$\nu_t=\tilde{\nu} f_{\nu1}$, where the $f_{\nu1}$ function is a damping function and is originally from Mellor and Herring \cite{mellor1968two}.
In addition to the SA model, Baldwin and Barth \cite{baldwin1990one,baldwin1991one}, an earlier work, also resorted to an auxiliary variable.
}
Furthermore, $\tilde{S}$ in Eq. \eqref{eq:SA} is given by $\tilde{S}=\Omega+\tilde{\nu}/(\kappa^2d^2)f_{\nu2}$ with $\Omega$ being the vorticity magnitude.
Like $f_{\nu1}$, $f_{\nu2}$ is also a damping function.
Both, $f_{\nu1}$ and $f_{\nu2}$ are functions of $\chi=\tilde{\nu}/\nu$.
Last, $d$ in Eq. \eqref{eq:SA} is the distance to the closest wall, $f_w$ is a function of $r=\tilde{\nu}/(\tilde{S}\kappa^2d^2)$, and $\sigma$, $c_{b1}$, $c_{b2}$, and $c_{w1}$ are constants.

Spalart \ and Allmaras progressively calibrated their model against a few free-shear flows, the log layer, and the viscous sublayer, in that specific order.
In the following, we summarize these calibrations.
The summary will shed light on the constraints we must preserve to preserve basic calibrations like the law of the wall.

\begin{figure}
\centering
\includegraphics[width=0.8\textwidth]{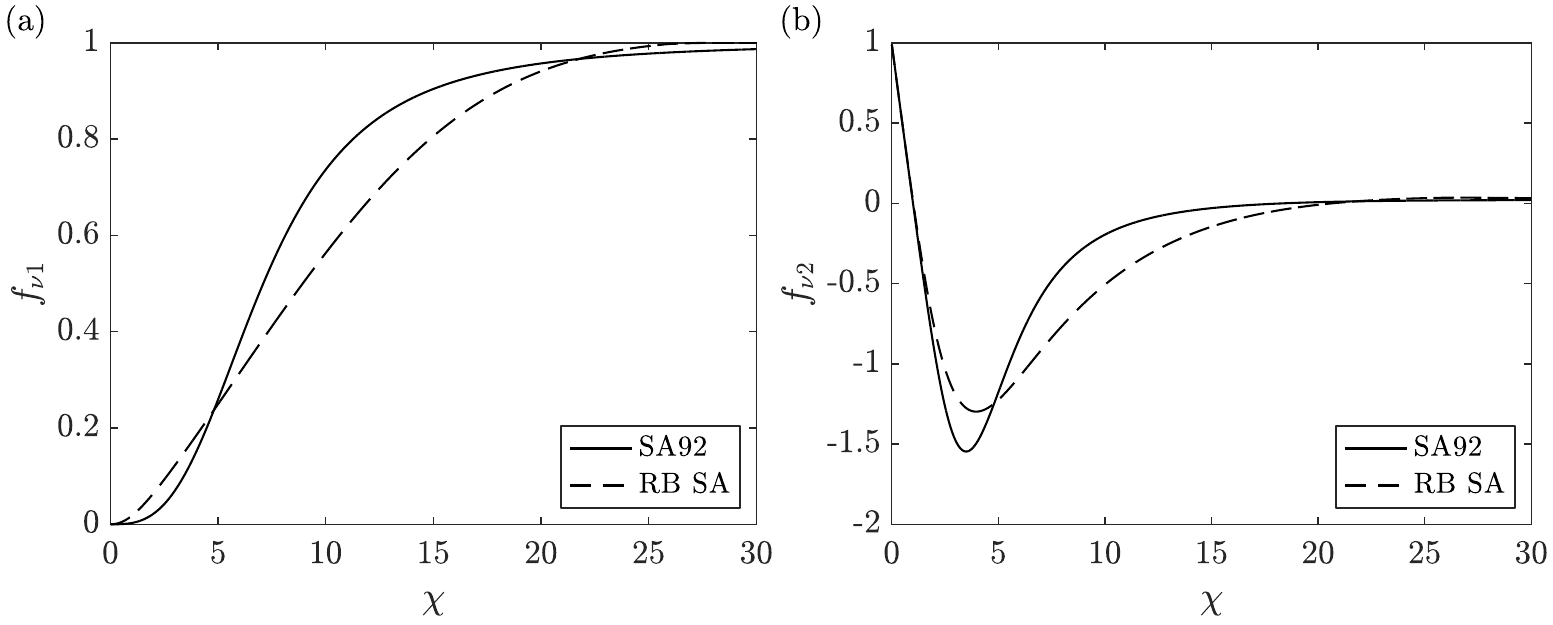}
\caption{Model function (a) $f_{\nu1}$ (b) $f_{\nu2}$ as a function of $\chi$.
}
\label{fig:fv}
\end{figure}

The calibration against the plane wakes and mixing layers led to two constraints for $c_{b1}$, $c_{b2}$, and $\sigma$, leaving one degree of freedom that was determined somewhat arbitrarily. 
In the log region, they set $\tilde{\nu}=\nu_t$ and $\tilde{S}=S$;
and considering that $U^+\sim\ln(y^+)/\kappa$ in the log layer, they set $f_{\nu1}=1$, $f_{\nu2}=0$, and $r=1$.
Consequently, 
\begin{equation}
c_{w1}=c_{b1}/\kappa^2+(1+c_{b2})/\sigma, ~\text{and}~~f_w(r=1)=1.
\label{eq:SA-log}
\end{equation}
In the viscous layer, $\tilde{\nu}=u_\tau \kappa d$ and $\tilde{S}=u_\tau/(\kappa d)$, and they had
\begin{equation}
f_{\nu1}=\frac{1}{\chi}\nu_{t,{\rm LoW}}^+,~~\text{and}~~f_{\nu2} = 1 - \frac{\chi}{1+\chi f_{\nu1}}.
\label{eq:SA-vis}
\end{equation}
Here, the eddy viscosity $\nu_{t,{\rm LoW}}^+$ is approximately
\begin{equation}
    \nu_t^+ \approx \kappa y^+ D,
\end{equation}
$D$ is the damping function
\begin{equation}
    D = \left( 1 - \exp \left( - \frac{y^+}{A} \right) \right)^2,~~A=17,
    \label{eq:vanD}
\end{equation}
$\chi = \kappa y^+$ in the viscous sublayer and the log layer.
Hence, 
\begin{equation}
    f_{\nu1} \approx \left( 1 - \exp \left( - \frac{\chi}{\kappa A} \right) \right)^2, ~~f_{\nu2}=1-\frac{\chi}{1+\chi [1-\exp(-\chi/(\kappa A))]^2}\label{eq:SA-fnu1}.
\end{equation}
Figure \ref{fig:fv} shows $f_{\nu1}$ and $f_{\nu2}$.
{\color{black}
The difference between the $f_{\nu1}$ and $f_{\nu2}$ here and those in Ref. \cite{spalart1992one} is mainly due to the use of the damping function in Eq. \eqref{eq:vanD}, which yields the right asymptotic behavior at the wall ($\nu_t\sim y^3$) and provides a closer agreement with the mean flow at high Reynolds numbers \cite{lee2015direct,bin2023data}.
}

If we further calibrate against the wake layer of a fully developed plane channel, we would have
\begin{equation}
    f_w = \frac{c_{b1}\dfrac{{\rm d} U_{\rm DNS}}{{\rm d} y}\nu_{t,{\rm DNS}} + 
    \dfrac{1}{\sigma}\left[\dfrac{{\rm d}}{{\rm d} y}\left( (\nu+\nu_{t,{\rm DNS}})\dfrac{{\rm d} \nu_{t,{\rm DNS}}}{{\rm d} y} \right) + c_{b2}\left(\dfrac{{\rm d} \nu_{t,{\rm DNS}}}{{\rm d} y}\right)^2\right]}{c_{w1}(\tilde{\nu}/d)^2}.
    \label{eq:SA-channel}
\end{equation}
The equation can be simplified 
\begin{equation}\label{eq:SA-fwLt1}
    f_w = \frac{c_{b1}}{\kappa^2 c_w} \frac{1}{r} + \frac{1+c_{b2}}{\sigma c_w} \frac{F_1^2}{F_0^2} + \frac{1}{\sigma c_w} \frac{F_2}{F_0^2}, 
\end{equation}
with
\begin{equation}
    F_0 =\frac{\nu^+_{t,{\rm DNS}}}{d^+},~~F_1 = \frac{{\rm d} \nu^+_{t,{\rm DNS}}}{{\rm d}  y^+},~~F_2 = (1 + \nu^+_{t,{\rm DNS}})\frac{{\rm d} ^2 \nu^+_{t,{\rm DNS}}}{{\rm d}  y^{+2}}.
\end{equation}
Here, {\color{black}
$F_0$, $F_0$, and $F_2$ can be obtained from DNS channel data \cite{lee2015direct} and are functions of $y$, or $r$.
}
In a fully-developed plane channel, $r$ takes values between approximately $0.4$ and 1.

We see that the terms, the constants, and the functions in the SA model are indeed coupled.
Consider, for example, $c_{b1}$.
The constant controls the model's behaviors in free-shear flows.
Meanwhile, $c_{b1}$ appears in Eqs. \eqref{eq:SA-log} and \eqref{eq:SA-channel}, and therefore varying its value will affect the calibration against the log layer, the viscous layer, and the wake layer of a plane channel.
The above argument applies to the other constants in the model as well.
We intend to always preserve the calibrations against the log layer, the viscous layer, and the wake layer of a plane channel flow.
To that end, we need to preserve the Eqs. \eqref{eq:SA-log}, \eqref{eq:SA-vis}, and \eqref{eq:SA-channel}.
The above is the basic idea of {\color{black}constrained re-calibration} or rubber band model.

\begin{figure}
\centering
\includegraphics[width=0.8\textwidth]{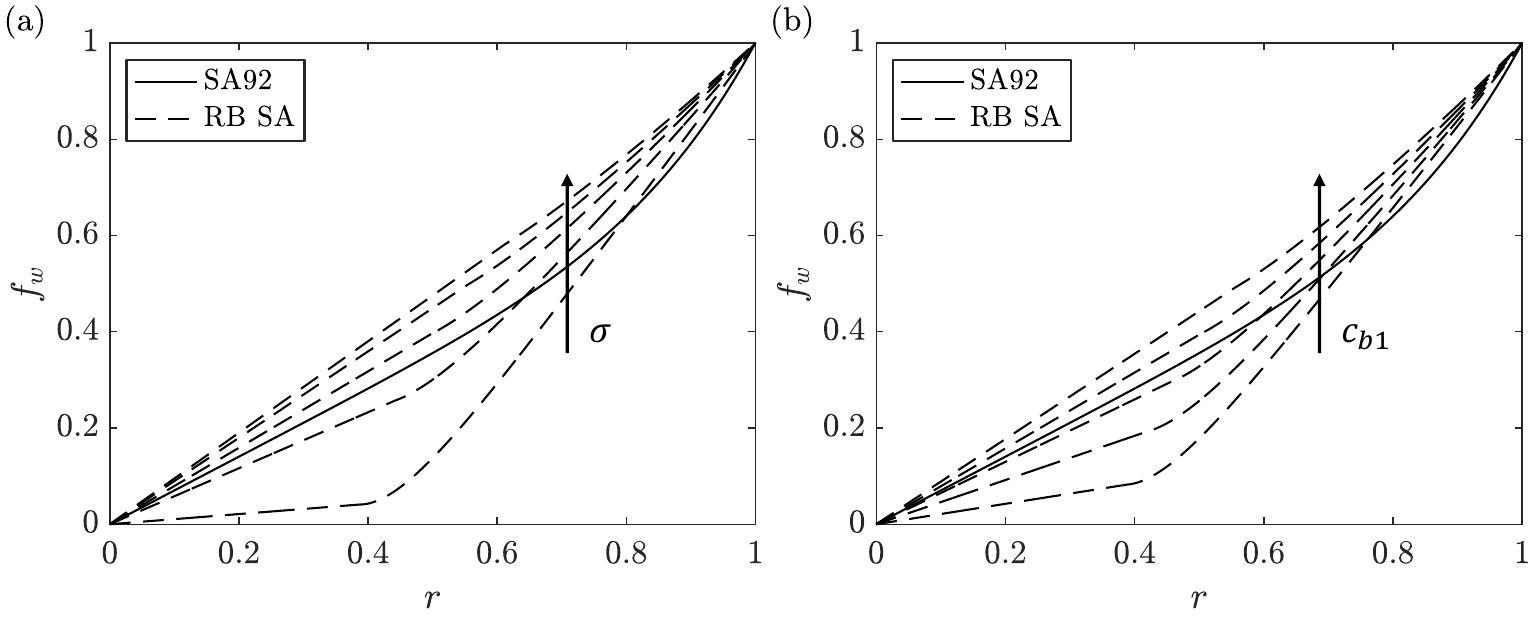}
\caption{$f_w$ as a function of $r$. (a) varying $\sigma$ (b) varying $c_{b1}$.
The baseline SA is recovered when $\sigma=2/3$ and $c_{b1}=0.1355$.}
\label{fig:fwFree}
\end{figure}

\begin{figure}
\centering
\includegraphics[width=0.8\textwidth]{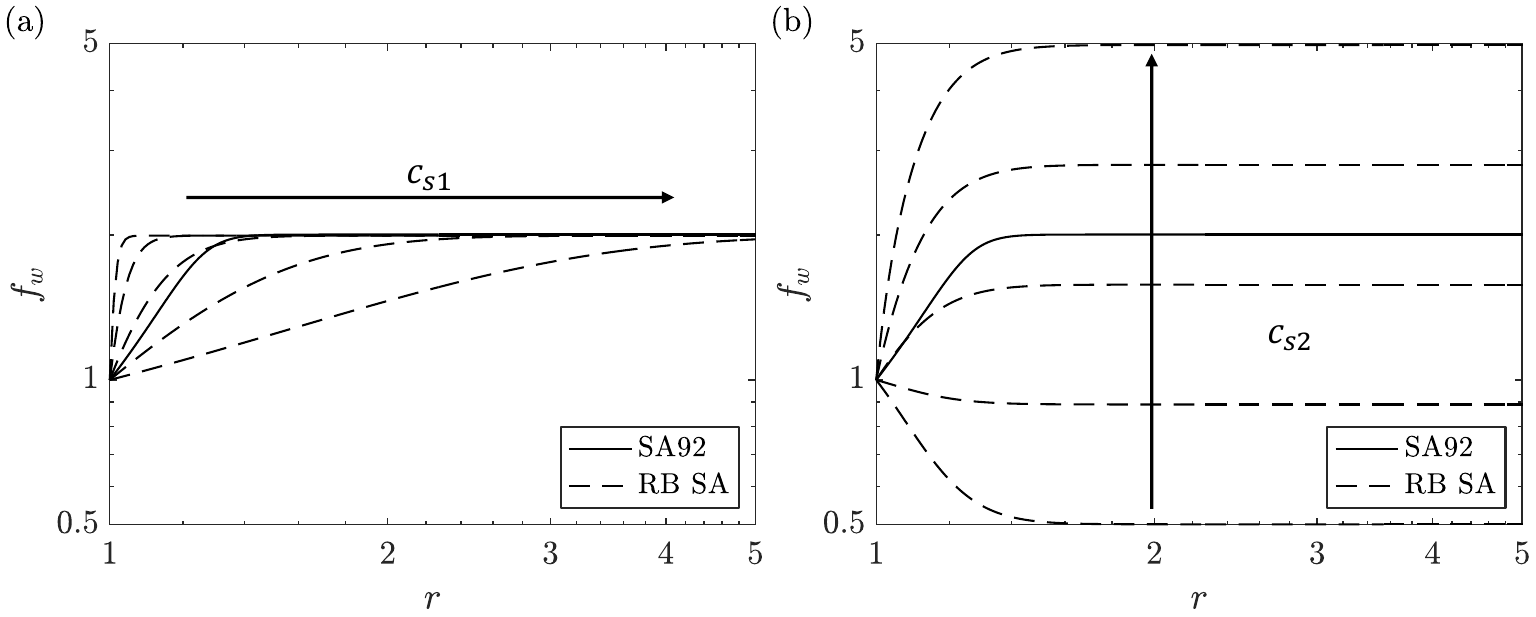}
\caption{$f_w$ as a function of $r$.(a) varying $c_{s1}$ (b) varying $c_{s2}$. 
The baseline SA is recovered when $c_{s1}=0.25$ and $c_{s2}=0.65$.
}
\label{fig:fwcs}
\end{figure}

In an effort to reduce the number of adjustable constants, we also keep the coefficient in front of the term $(\partial \tilde{\nu}/\partial x_i)^2$, the same as that in the baseline SA. 
This leads to 
\begin{equation}
    c_{b2}=2.433\sigma-1.
    \label{eq:SA-cb2}
\end{equation}
Furthermore, we require that $f_w$ between $r=0$ and $r\approx 0.4$ be a linear function of $r$.
{\color{black}The behavior of $f_w$ for small $r$'s is likely not going to have an impact on the model's behavior, and therefore more sophisticated functional forms are not pursued.}
With these constraints, we may freely specify $c_{b1}$, $\sigma$, and the value of $f_w$ for $r>1$.
For simplicity, we parameterize $f_w$ for $r>1$ as follows
\begin{equation}
f_w=A\tanh\left(\frac{r-1}{B}\right)+1,~~r>1,
\label{eq:fwRGt1}
\end{equation}
where
\begin{equation}
A=10^{2c_{s2}-1}-1,~~B= 10^{4c_{s1}-1}/5,
\label{eq:AB}
\end{equation}
$c_{s1}$ and $c_{s2}$ are two adjustable constants, $c_{s2}$ is the value of $f_w$ at large $r$ values, $c_{s1}$ controls how quickly $f_w$ gets to $c_{s2}$ as $r$ increases.
{\color{black}
Figure \ref{fig:fwFree} shows $f_w$ for $r<1$ for a couple randomly picked $c_{b1}$ and $\sigma$ values.
}
Figure \ref{fig:fwcs} shows $f_w$ for $r>1$ as we vary $c_{s1}$ and $c_{s2}$. 
Equations \eqref{eq:SA}, \eqref{eq:SA-log}, \eqref{eq:SA-fnu1}, \eqref{eq:SA-channel}, \eqref{eq:SA-cb2}, \eqref{eq:fwRGt1}, and \eqref{eq:AB} constitute our re-calibration: 
{\color{black}
Eq. \eqref{eq:SA} is the transport equation for $\tilde{\nu}$, Eqs. \eqref{eq:SA-log}, \eqref{eq:SA-fnu1}, and \eqref{eq:SA-channel} are constraints that preserve the law of the wall and the profile in the wake layer in a channel, and \eqref{eq:AB} is an additional constraint to limit the number of adjustable constants.
}
{\color{black}
The previous machine learning methods like FIML do not necessarily preserve the constraints in Eqs. \eqref{eq:SA-log}, \eqref{eq:SA-fnu1}, and \eqref{eq:SA-channel}, and therefore the resulting models do not preserve the behavior of the baseline model in a channel and a flat plat.
Here, by ensuring the preservation of the law of the wall, the hope is that the resulting model has better generalizability.
}

\subsection{Re-calibration}
\label{sub:bayesian}

Given training data and a few degrees of freedom in the RANS model, the objective of model re-calibration is to adjust the few degrees of freedom such that the trained model captures the training data.
Formally, the problem of model re-calibration is 
\begin{equation}
\min_{\mathbf{x}} \mathcal{E}(\mathbf{x}).
\label{eq:BO-setup}
\end{equation}
{\color{black}
Here, $\mathcal{E}$ is the error in the model prediction.
Specifically,
\begin{equation}
\mathcal{E}(\mathbf{x}) = \sum_{i=1}^{N} w_i\left(f_i(\mathbf{x}) - f_{i,r}\right)^2,
\end{equation}
where $N$ is the number of reference data, $f_{i,r}$ is the reference data, $f_i$ is the model prediction, $w_i\equiv 1$ is a weight.
In the following section, we will study round jets, plane jets, airfoil, wall-mounted hump, and prolate spheroid.
For the jets, the reference data $f_i$ is the spreading rate obtained experimentally \cite{witze1976turbulent,bradbury1965structure}.
For airfoils, the reference data $f_i$ is the lift coefficient obtained from experimental measurements \cite{coles1979flying}.
For wall-mounted hump, the reference data $f_i$ is the skin friction coefficient at two axial locations, which are also experimental measurements \cite{seifert2002active}.
Lastly, for the prolate spheroid, the reference data $f_i$ is the pressure coefficient obtained experimentally \cite{chesnakas1997detailed,wetzel1998measurement}.
}
The bold variable $\mathbf{x}$ in Eq. \eqref{eq:BO-setup} represents the degrees of freedom in the model. 
The present paper focuses on identifying a good ${\bf x}$ space.
The minimization problem can be solved by invoking any existing machine-learning method.
Here, we resort to Bayesian optimization (BO) for its simplicity.
{\color{black} We shall show that by addressing the {\it where} aspect, even simple methods like BO would yield generalizable models.}

The procedure of BO is as follows.
First, we model the error function $\mathcal{E}'$ as a Gaussian process. 
Consider $n$ samples in the design space, denoted as $\mathbf{X}^* = \left[\mathbf{x}^*_1, \mathbf{x}^*_2, \cdots, \mathbf{x}^*_n\right]^T \in \mathbb{R}^{n \times d}$, where $d$ is the dimension of the parameter space. 
The error function is given by
\begin{equation}
\mathcal{E}'(\mathbf{X}) \sim \mathcal{N}(M(\mathbf{X}),\Sigma(\mathbf{X})),
\end{equation}
where $\mathbf{X}\in \mathbb{R}^{m \times d}$ are where an estimate of $\mathcal{E}$ is needed, $M(\mathbf{X})\in \mathbb{R}^{m \times 1}$ is the mean vector, and $\Sigma(\mathbf{X})\in \mathbb{R}^{m\times m}$ is the covariance matrix.
The mean vector and covariance matrix are given by
\begin{align}
M(\mathbf{X}) &= K(\mathbf{X},\mathbf{X}^*)K(\mathbf{X}^*,\mathbf{X}^*)^{-1}\mathcal{E}(\mathbf{X}^*),\\
\Sigma(\mathbf{X})&=K(\mathbf{X},\mathbf{X}) - K(\mathbf{X},\mathbf{X}^*)K(\mathbf{X}^*,\mathbf{X}^*)^{-1}K(\mathbf{X}^*,\mathbf{X}),
\end{align}
where
\begin{equation}
    K(\mathbf{X},\mathbf{X})_{ij} = \exp\left(-\frac{\left(\mathbf{x}_i - \mathbf{x}_j\right)^T\left(\mathbf{x}_i - \mathbf{x}_j\right)}{2}\right).
\end{equation}
This estimate gives
\begin{equation}
\mathcal{E}'(\mathbf{x}_i) = \mu(\mathbf{x}_i) + \sigma(\mathbf{x}_i)z,
\end{equation}
at any $x_i$,
where $\mu(\mathbf{x}_i) = M_i$ is the mean and $\sigma(\mathbf{x}_i) = \Sigma_{ii}$ is the standard variation. 
The random number $z$ is drawn from a normal distribution.
Next, we compute the acquisition function.
Here, we employ the expected improvement acquisition function, defined as
\begin{align}
a(\mathbf{x}_i,\mathcal{E}') &= {\rm EI}\left[ \max\left( \mathcal{E}'(\mathbf{x}^*_j) - \mathcal{E}'(\mathbf{x}_i),0 \right) \right]\\
&= \sigma \left(\phi(z_0) + z_0 \int^{z_0}_{-\infty} \phi(z) {\rm d}z \right),
\end{align}
Here, $\rm EI$ stands for expected improvement, $z_0 = (\mathcal{E}'(\mathbf{x}^*_j) - \mu(\mathbf{x}_i))/\sigma$,
$\mathbf{x}^*_j$ corresponds to the $j$th observed point that minimizes the estimated error function $\mathcal{E}'$,
and $\phi$ is the standard normal distribution function.
The location where the acquisition function attains its maximum is identified as the next point to sample.
The above two steps are repeated until the process converges.
We will further illustrate this process in Section \ref{sub:jet}.

\subsection{Physical implications}
\label{sub:physics}

We can readily re-calibrate/augment the baseline SA following the steps highlighted in the two subsections above.
However, by relying solely on data, one would not get interpretable results.
In order to interpret the results, we need to develop a physical understanding of adjustable constants.

\begin{figure}
\centering
\includegraphics[width=0.8\textwidth]{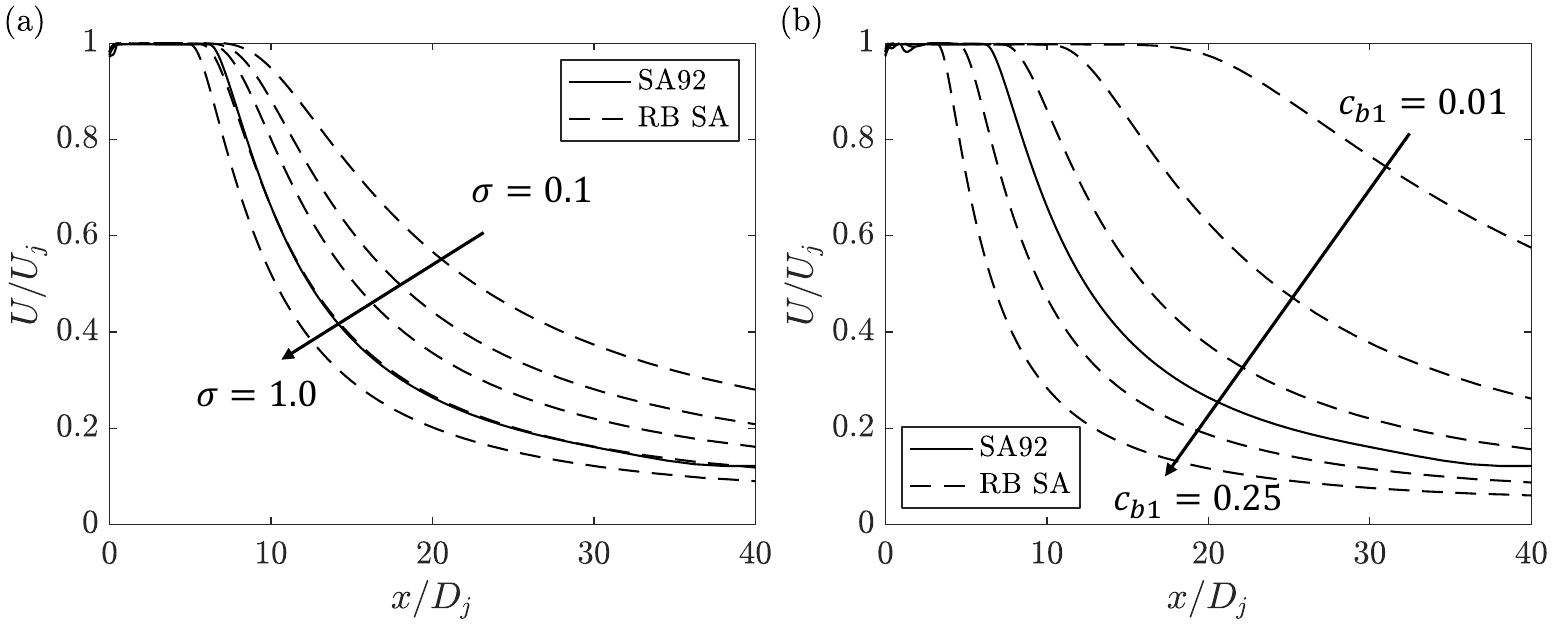}
\caption{Centerline velocity in an axis-symmetric jet. Here, $U_j$ is the jet velocity, and $D_j$ is the diameter of the jet. }
\label{fig:jetConEff}
\end{figure}

The two adjustable constants $c_{b1}$ and $\sigma$ are coefficients in front of the production term and the diffusion term and therefore control the model's behaviors in unbounded flows.
Increasing or decreasing $c_{b1}$ results in an increase or decrease, respectively, of the production term in the $\tilde{\nu}$ equation, making the flow more or less dissipative. 
Increasing or decreasing $\sigma$ leads to an increase or decrease, respectively, in the diffusion term in the $\tilde{\nu}$ equation, affecting the redistribution of the eddy viscosity in the flow.
To study the effects of $c_{b1}$ and $\sigma$, we study the model's behaviors in an axis-symmetric jet.
The flow is canonical and its configuration is detailed on the NASA Turbulence Modeling Resource (TMR) site (https://turbmodels.larc.nasa.gov)  and is not repeated here for brevity.
Figure \ref{fig:jetConEff} shows the centerline velocity as a function of the longitudinal coordinate as we vary $\sigma$ and $c_{b1}$.
Figure \ref{fig:jetConEff}(a) shows the result for $c_{b1}=0.1355$ (the value in the baseline SA) while we vary $\sigma$ between $0.1$ and $1.0$, and figure \ref{fig:jetConEff}(b) shows the results for $\sigma=2/3$ while we vary $c_{b1}$ between $0.01$ and $0.25$.
The flow becomes increasingly more diffusive as $\sigma$ increases, leading to faster-decaying jet and centerline velocity, as shown in figure \ref{fig:jetConEff}(a).
The flow becomes increasingly more dissipative as $c_{b1}$ increases, as shown in figure \ref{fig:jetConEff}(b).

\begin{figure}
\centering
\includegraphics[width=0.8\textwidth]{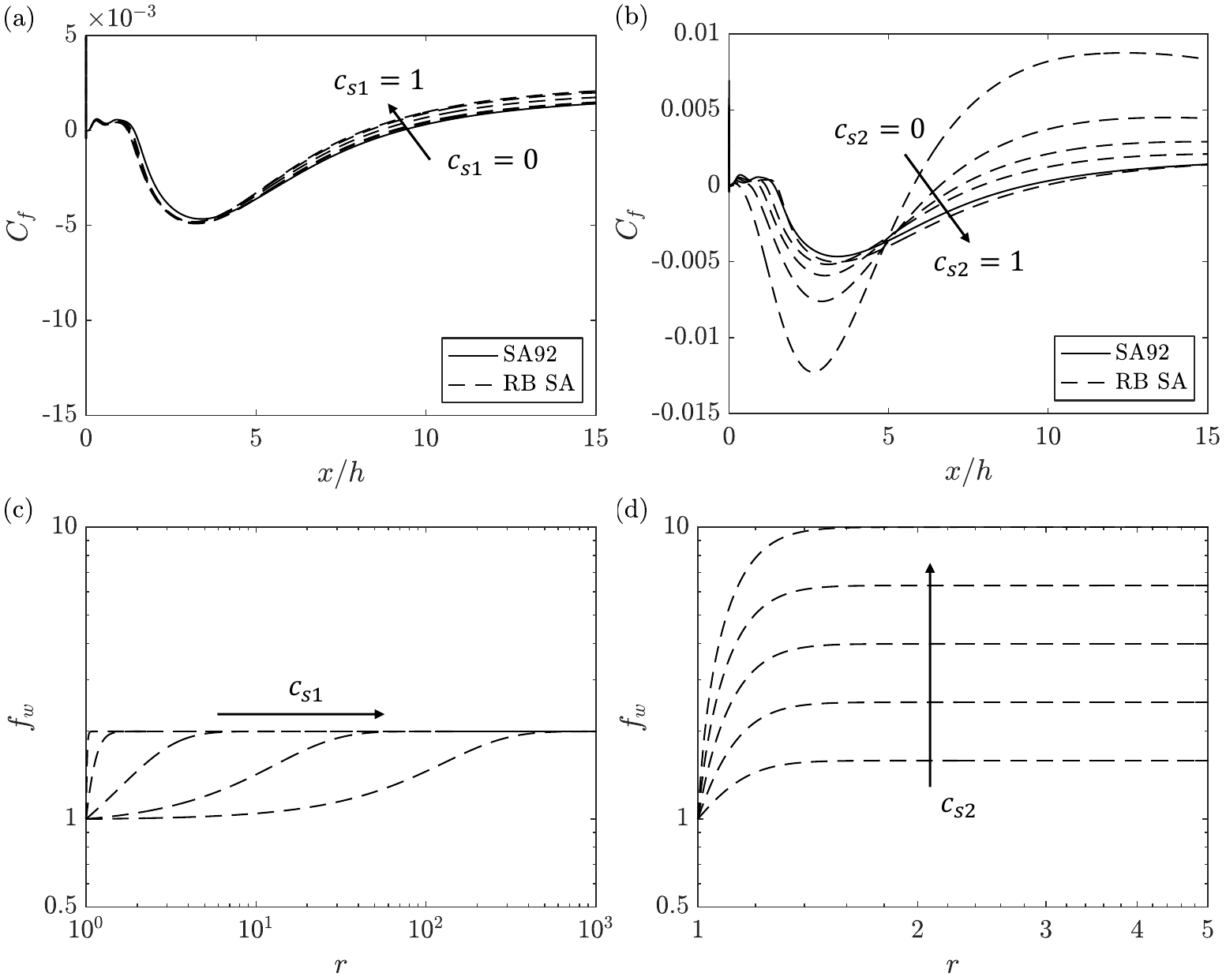}
\caption{Skin friction coefficient $C_f$ in the back-facing-step case when varying (a) $c_{s1}$ (b) $c_{s2}$ and (c,d) the
corresponding $f_w$.
Here, $h$ is the height of the step.}
\label{fig:bfsConEff}
\end{figure}

The adjustable constants $c_{s1}$ and $c_{s2}$ control $f_w$ and in turn the model's behaviors in non-equilibrium boundary-layer flows. 
Increasing or decreasing $c_{s1}$ and $c_{s2}$ leads to an increase or decrease, respectively, in the destruction term, making the model less or more dissipative.
The behavior of the baseline SA is recovered by setting $c_{s1}=0.65$ and $c_{s2}=0.25$.
To study the effects of $c_{s1}$ and $c_{s2}$, we study the model's behaviors in the back-facing step (BFS) flow.
The flow is canonical.
Its configuration could be found on the TMR site and therefore is not detailed here for brevity.
Figure \ref{fig:bfsConEff} shows the skin friction coefficient $C_f$. 
Figure \ref{fig:bfsConEff} (a) shows $C_f$ as we vary $c_{s1}$ from $0$ to $1$, with $c_{s2}$ held at a constant value of $0.65$. 
Figure \ref{fig:bfsConEff} (b) shows $C_f$ as $c_{s1}$ is held constant at $0.25$, while $c_{s2}$ is varied from $0.6$ to $1$.
The corresponding $f_w$ when varying $c_{s1}$ and $c_{s2}$ are plotted in figures \ref{fig:bfsConEff} (c,d), respectively.
We see that $c_{s1}$ controls the recovery, and $c_{s2}$ controls the strength of the recirculation.

\begin{figure}
\centering
\includegraphics[width=0.45\textwidth]{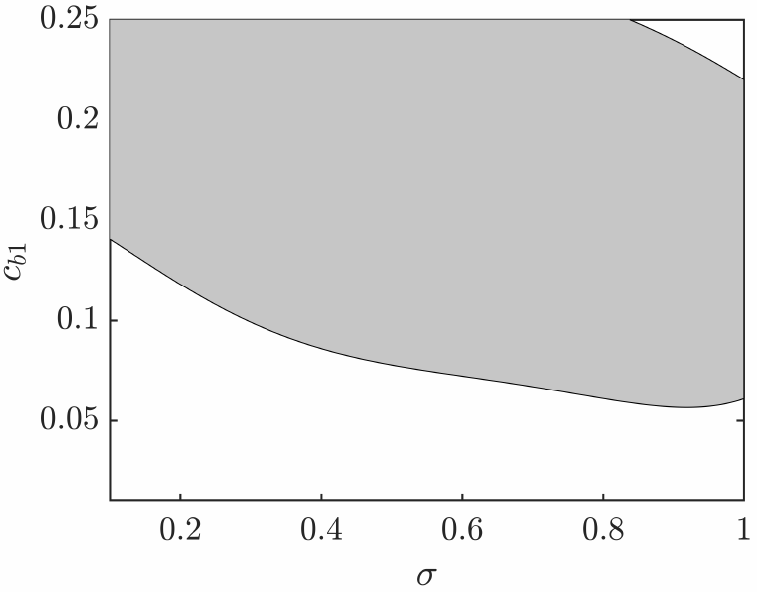}
\caption{The recommended range of $c_{b1}$ and $\sigma$.}
\label{fig:2range}
\end{figure}

Table \ref{tab:constantRange} gives the ranges of the four adjustable constants.
In anticipation of the basic calibration against a flat-plate boundary layer, we constrain the value of $c_{b1}$ and $\sigma$ as shown in figure \ref{fig:2range}.
Larger ranges are unnecessary because the present ranges already offer good versatility, as we will see shortly.
Lastly, we note that the four adjustable constants can potentially vary locally as a function of local fluid variables, although, in this paper, we will treat them as global parameters.
{\color{black}This is again a simplification.
A number of the recent machine learning frameworks allow the coefficients to vary within the flow.
By not employing these more sophisticated methods, our objective is to show that by addressing the {\it where} aspect of the problem, simple methods like BO yield generalizable models.
}

\begin{table}
\caption{\label{tab:constantRange} The ranges of four adjustable constants.}
\centering
\begin{minipage}{\textwidth}
\centering
\begin{tabular}{cc}
\hline
Constant & Range\\
\hline
$\sigma$ & $[0.1,1.0]$\\
$c_{b1}$ & $[0.01,0.25]$\\
$c_{s1}$ & $[0,1]$\\
$c_{s2}$ & $[0,1]$\\
\hline
\end{tabular}
\end{minipage}
\end{table}

\section{Results}
\label{sec:results}

In this section, we recalibrate the SA according to our reformulation to tackle some long-standing issues in RANS modeling, including the plane-jet/round-jet anomaly, airfoil stall, recovery after separation, and secondary flow separation.
Re-calibration relies on Bayesian optimization in the 4D space given in Table \ref{tab:constantRange}. 
The objective is twofold.
First, we aim to show that these identified degrees of freedom allow us to ``stretch'' the baseline model to accommodate specific needs.
This is something all machine learning methods should be able to do.
Second, we aim to show that adjusting these identified constants is not detrimental to the baseline model---this is missing and is the focus of this study.

\subsection{Code and RANS setup}

We employ the open-source second-order finite-volume software OpenFOAM V2106 for all our calculations.
Details of the code can be found in Ref. \cite{jasak2007openfoam} and are not repeated here for brevity.
The detailed implementation of the reformulated SA model is available on GitHub \cite{GitHubLink}.
The flows we consider include fully-developed channel, flat-plate boundary layer, round jet, plane jet, wall-mounted hump, NACA4412 airfoil at angles of attack prior- and post-stall, and 6:1 prolate spheroid.
Detailed setup of all cases, except for the 6:1 prolate spheroid case, are available on the TMR site, including the mesh, the inflow condition, and the boundary conditions. 
The setup of the 6:1 prolate spheroid is detailed in Ref. \cite{xiao2007prediction,amiri2019rans}.

\subsection{Channel and Flat-plate Boundary Layer} 
\label{sec: validation}

We first verify that the four adjustable constants have little to no effect on fully developed plane channel and flat-plate boundary layer.
To that end, we randomly vary the adjustable constants within their respective ranges as shown in Table \ref{tab:constantRange}. 
The other constants, i.e., $c_{b2}$, and functions, $f_{\nu1}$, $f_{\nu2}$, and $f_w(r<1)$, vary according to Eqs.\eqref{eq:SA-log}, \eqref{eq:SA-vis}, \eqref{eq:SA-channel}, \eqref{eq:SA-cb2}, and \eqref{eq:AB}.
We denote these results as ``constrained re-calibration.''
For comparison purposes, we vary $c_{b1}$ and $\sigma$ while keeping $c_{b2}$, $f_{\nu1}$, $f_{\nu2}$, and $f_w$ unchanged.
We denote these results as ``unconstrained re-calibration.''

\begin{figure}
\centering
\includegraphics[width=0.8\textwidth]{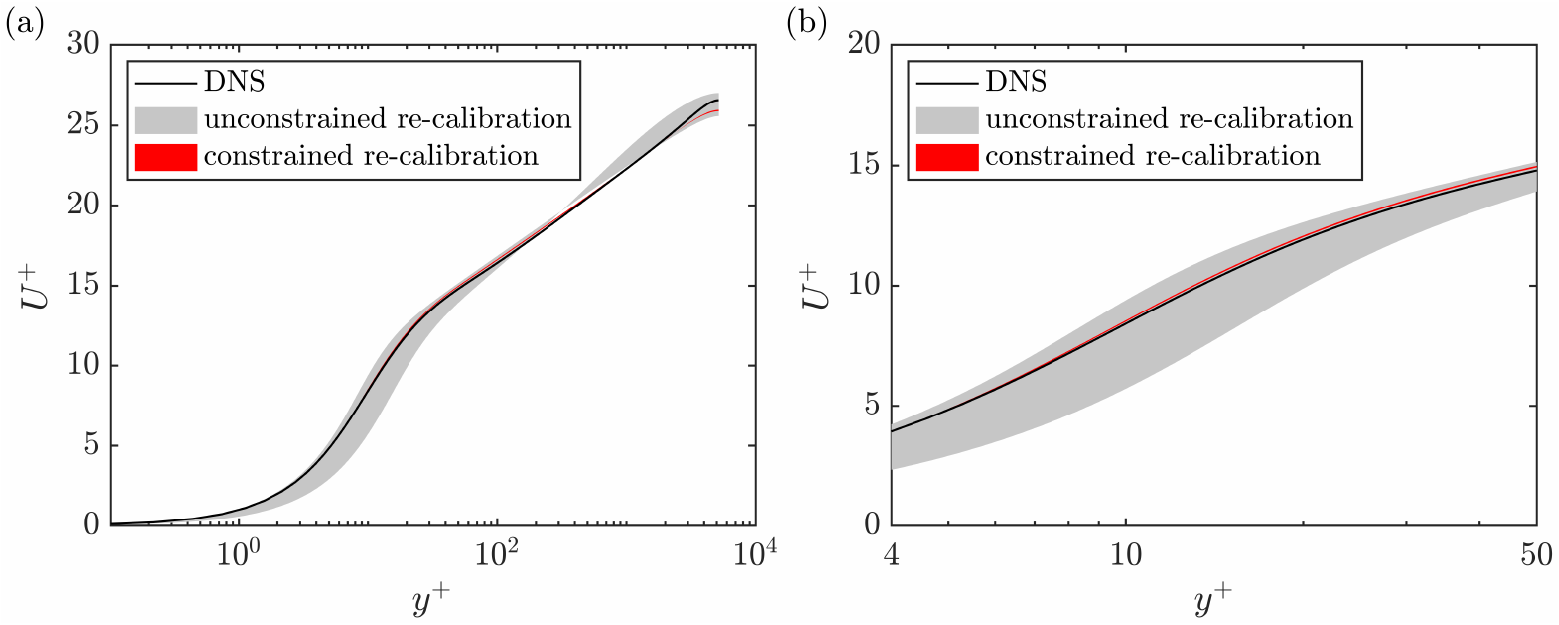}
\caption{(a) Mean velocity $U^+$ as a function of $y^+$ in a $Re_{\tau} = 5200$ fully developed channel. 
(b) Same as (a), but a zoom-in to the buffer layer.}
\label{fig:chanVal}
\end{figure}

Figure \ref{fig:chanVal} shows the channel flow results at $Re_\tau=5200$.
Figure \ref{fig:chanVal} (a) shows the velocity profile across the half channel, and figure \ref{fig:chanVal} (b) zooms into the buffer layer.
We show the DNS result for comparison purposes.
The constrained re-calibration results are confined to the red region, and the grey region shows the variation of the unconstrained re-calibration results.
We observe the following.
The constrained re-calibration results follow the DNS result irrespective of the four adjustable constants.
On the other hand, unconstrained re-calibration does not preserve the channel flow calibration as we vary the constants.
Specifically, significant variations in the SA's prediction are found in the viscous layer and the wake layer.
It is worth noting that there is a small discrepancy between the baseline SA and the DNS near the channel centerline.
This discrepancy can be removed by employing a different set of basic calibrations \cite{bin2023data}, which is out of the scope of this paper.

\begin{figure}
\centering
\includegraphics[width=0.4\textwidth]{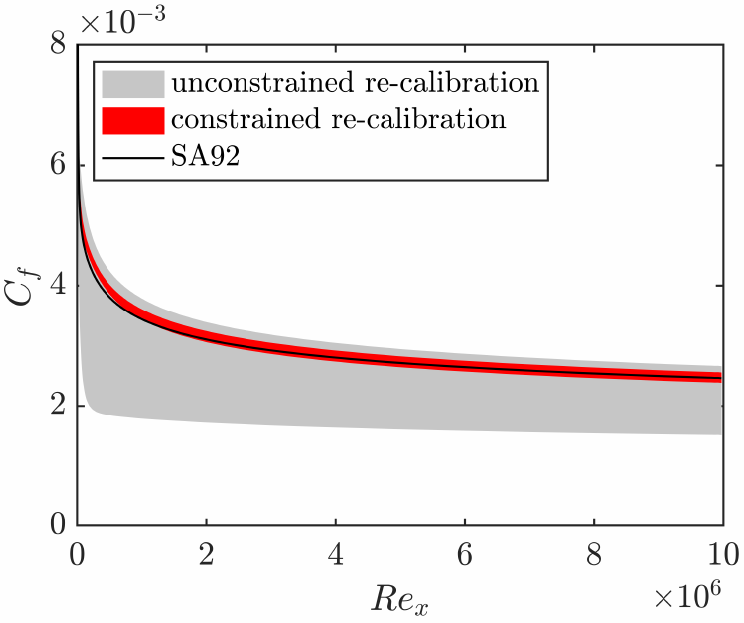}
\caption{Skin friction coefficient of a flat-plate boundary layer.}
\label{fig:validation}
\end{figure}

Figure \ref{fig:validation} shows the skin friction coefficient as a function of $Re_x \equiv U_{\rm inf} x/\nu$ for a flat-plate boundary layer, where $U_{\rm inf}$ is the freestream velocity.
The prediction of the baseline SA model is shown for reference purposes.
The red region confines the constrained re-calibration result, and the grey region shows the variation of unconstrained re-calibration.
We see that the constrained re-calibration result follows the baseline SA regardless of the adjustable constant, but adjusting the constants has a significant impact on the baseline SA's result.
The slight variation of the constrained re-calibration result as we vary the adjustable constant is due to the defect layer not being part of the basic calibration. 

We have shown that constrained re-calibration preserves the basic calibrations against fully developed channel and zero-pressure-gradient flat-plate boundary-layer flows.
In the following subsections, we will not repeat these results, with the understanding that the adjustable constants do not affect the channel and boundary-layer flows.

\begin{figure}
\centering
\includegraphics[width=0.4\textwidth]{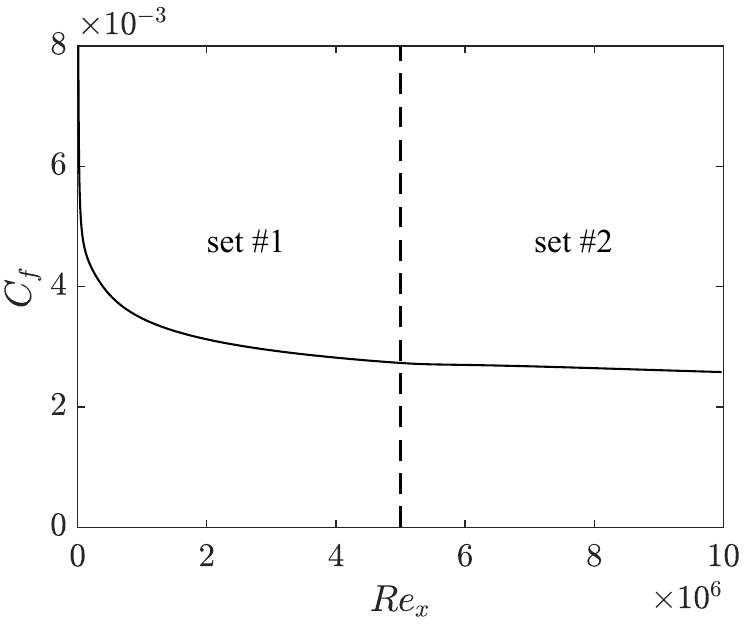}
\caption{The skin friction coefficient $C_f$ as a function of $Re_x$, where two arbitrarily picked sets of model constants are used.}
\label{fig:cfDiff}
\end{figure}

\begin{figure}
\centering
\includegraphics[width=0.4\textwidth]{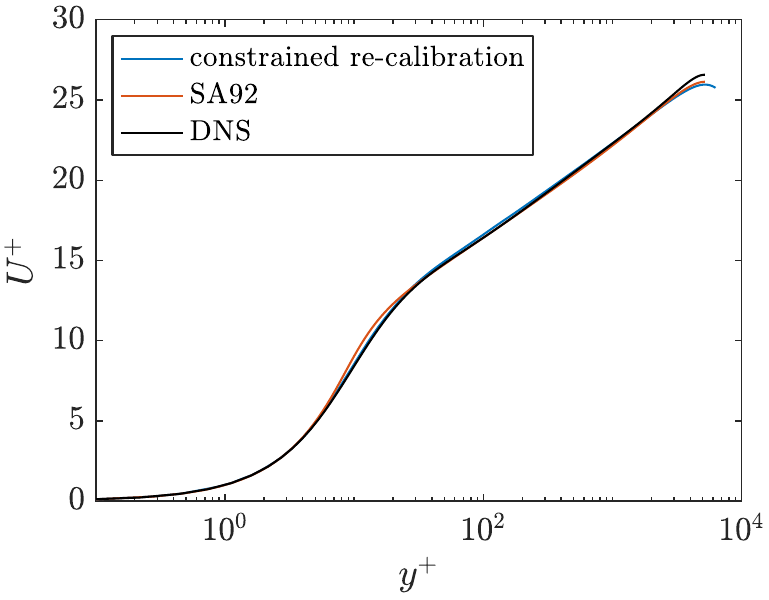}
\caption{Channel profile. The adjustable constants are varied randomly in the flow field.}
\label{fig:channelChangeConResult}
\end{figure}

Before we proceed, we consider the following numerical experiment, where we employ two arbitrary sets of model constants upstream and downstream of $Re_x=5\times 10^6$. 
Figure \ref{fig:cfDiff} shows the result.
We see that abruptly changing the model constants within the domain does not generate any numerical shock in OpenFOAM.
{\color{black}
Furthermore, we vary $\sigma$, $c_{b1}$, $c_{s1}$, and $c_{s2}$ randomly in a plane channel.
Figure \ref{fig:channelChangeConResult} shows the resulting velocity profile.
We see that varying the four adjustable constants randomly has no impact on the channel flow results.
The above is a desirable feature, as the adjustable constants in a RB model can potentially vary locally, providing a good playground for existing machine learning paradigms like FIML.}

\subsection{Round-jet/plane-jet Anomaly} 
\label{sub:jet}

The spreading rate of a jet is defined as follows
\begin{equation}
S=\frac{{\rm d}l_{1/2}}{{\rm d}x},
\label{eq:S}
\end{equation}
where $l_{1/2}$ is the distance from the jet centerline to the location where the jet velocity decays to half its centerline value. 
Both round and plane jets are self-similar sufficiently downstream, where $S$, as defined in Eq.\eqref{eq:S}, is a constant.
The round-jet/plane-jet anomaly refers to the mismatch between the model-predicted spreading rate and the experimental measurements \cite{pope1978explanation,witze1976turbulent,bradbury1965structure}:
when a model is calibrated against a plane jet, it would mispredict the growth rate of a round jet, and vice versa.
Most RANS models suffer from this round-jet/plane-jet anomaly, including the $k$-$\epsilon$, $k$-$\omega$, $k$-$\omega$ SST, and SA92 \cite{bardina1997turbulence}.
Here, we attempt to resolve this issue for the SA model.
The SA92 model captures the spreading rate of the plane jet but fails to predict the spreading rate of the round jet \cite{bridges2010establishing,bridges2011nasa}. 
To address this anomaly, we need to re-calibrate the model behaviors in unbounded flows.
The constants $c_{s1}$ and $c_{s2}$ have negligible effects on free-shear flow, thus addressing the around-jet/plane-jet anomaly requires re-calibrating $c_{b1}$ and $\sigma$. 

\begin{figure}
\centering
\includegraphics[width=\textwidth]{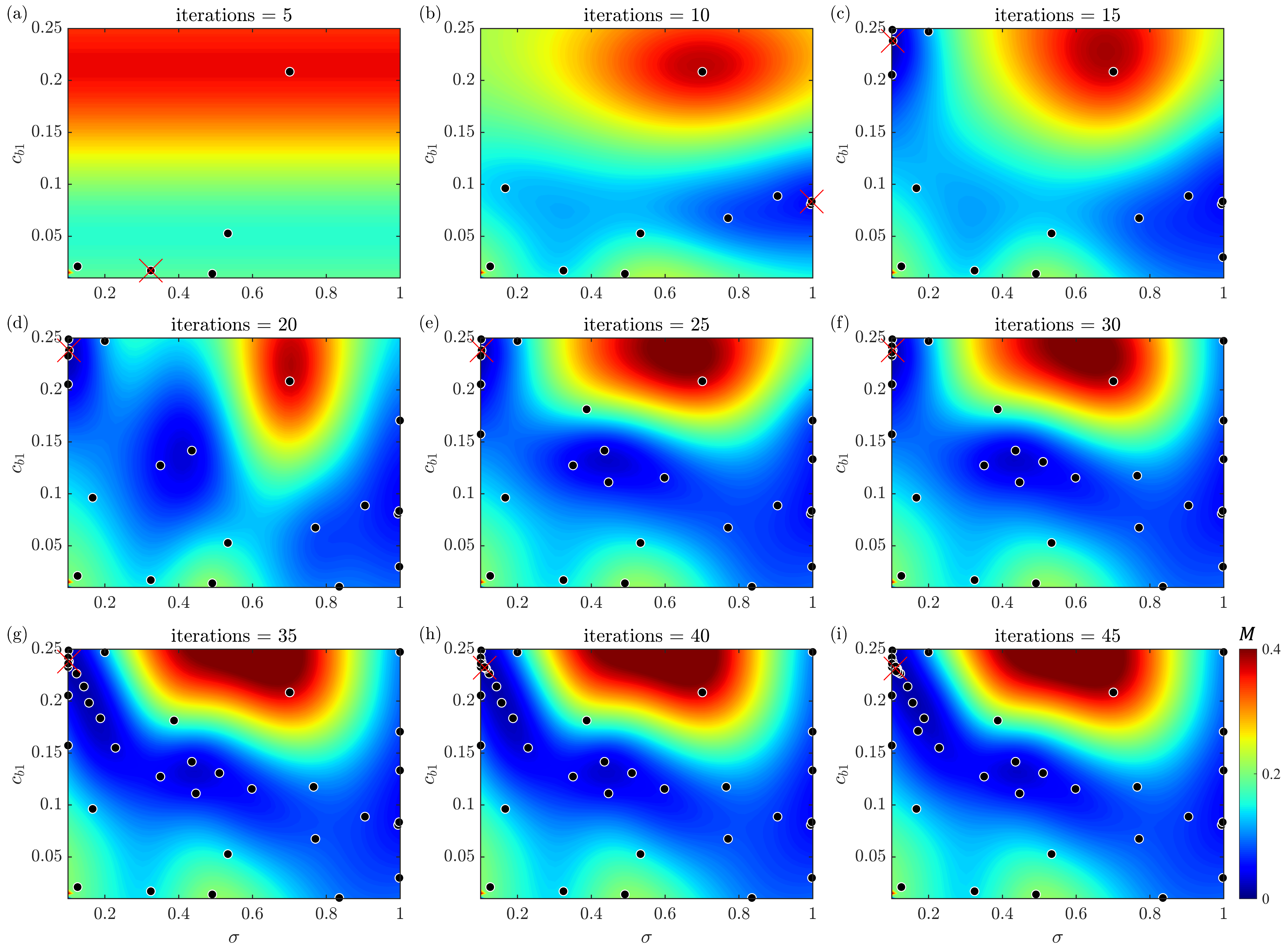}
\caption{
Contours of the estimated error function $M$ as a function of $\sigma$ and $c_{b1}$. (a-i) as iteration increases. The optimum at a given iteration is marked using a red cross symbol.
The sample locations are highlighted using black circle symbols.
}
\label{fig:bayeMu}
\end{figure}

\begin{figure}
\centering
\includegraphics[width=0.4\textwidth]{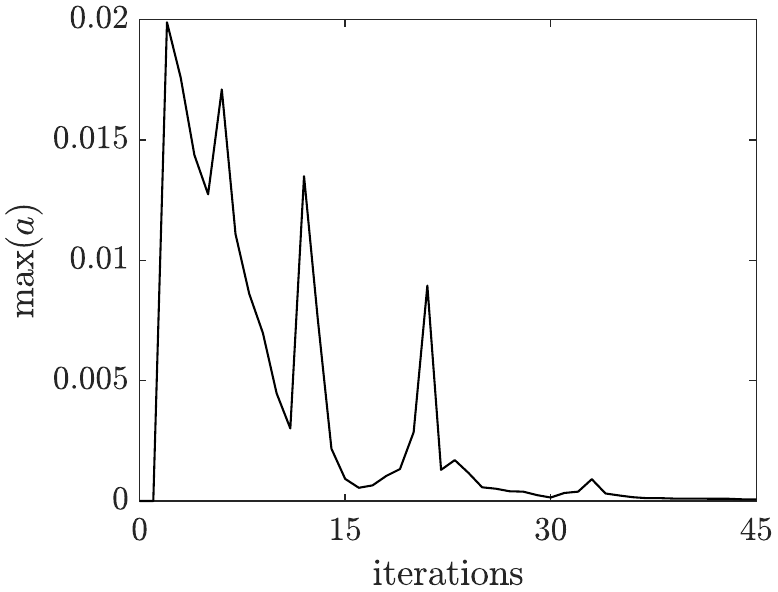}
\caption{
The maximum of the acquisition function as a function of the iteration.
}
\label{fig:bayeA}
\end{figure}

We first illustrate the Bo-based re-calibration process. 
Figure \ref{fig:bayeMu} shows the evolution of the estimated error function $M$ as a function of $\sigma$ and $c_{b1}$ as iteration increases.
We see that the optimization converges in about 25 iterations, after which the error function barely changes, and the subsequent samples are located in the neighborhood of the optimum. 
Figure \ref{fig:bayeA} shows the maximum value of the acquisition function as a function of the iteration.
We see that it is generally a decreasing function of the iteration and is very close to 0 after 25 iterations.
This suggests the expected improvement beyond the known optimum is small after 25 iterations.
This process gives rise to an optimum located at $\sigma=0.1$ and $c_{b1}=0.24$. 
In contrast, the baseline SA model's constants, $\sigma=0.667$ and $c_{b1}=0.1355$.


\begin{figure}
\centering
\includegraphics[width=0.8\textwidth]{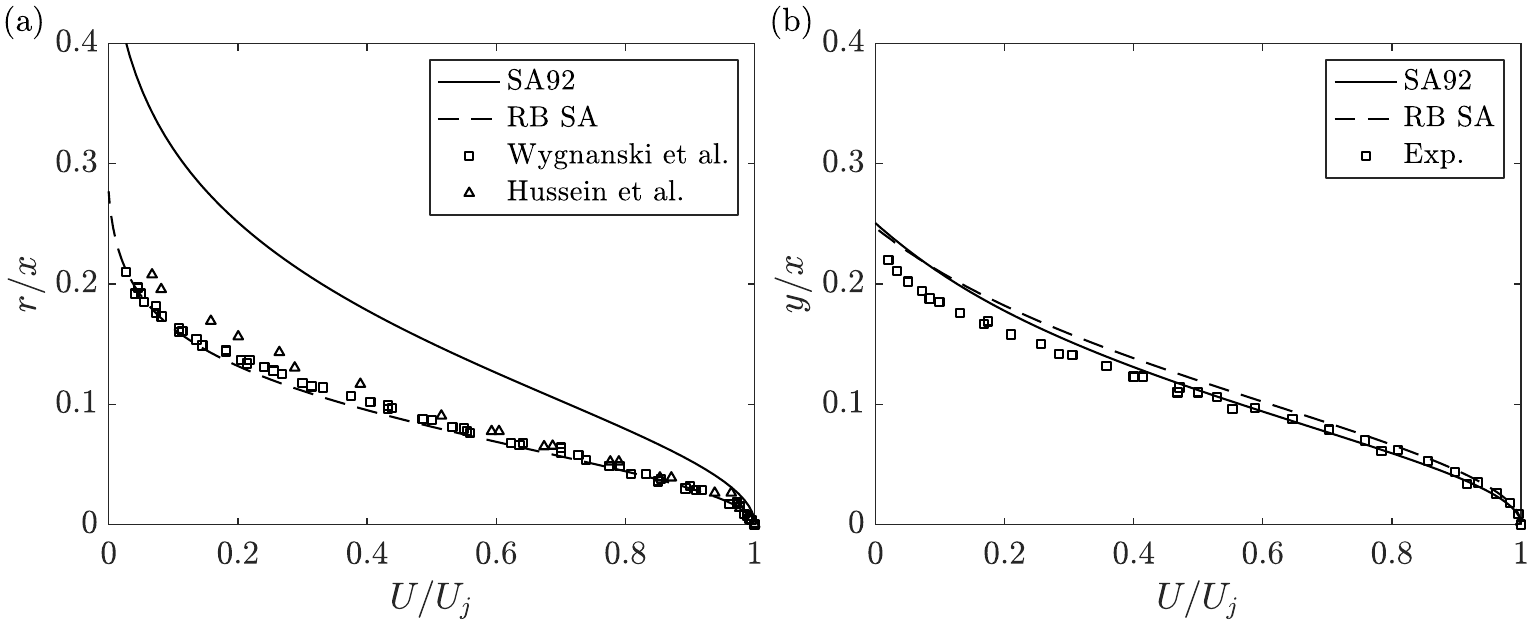}
\caption{
Velocity profiles in the self-similar regime.
(a) round jet (b) plane jet.
The experimental data are from \cite{wygnanski1969some,hussein1994velocity, bradbury1965structure}.
Here, $r$ and $y$ are transverse coordinates for the round jet and plane jet.
$x$ is the streamwise coordinate.
}
\label{fig:jetAno}
\end{figure}

Figure \ref{fig:jetAno} shows the velocity profiles as a function of the transverse coordinate for both round and plane jets in the self-similar region.
The results due to constrained re-calibration are referred to as ``RB SA''.
{\color{black}
The prediction of the baseline SA is included for reference.
In Fig. \ref{fig:jetAno}, we have included experimental measurements reported in Ref. \cite{wygnanski1969some}, Ref. \cite{hussein1994velocity} and Ref. \cite{bradbury1965structure}.
We see that the early experiment and the more recent ones give essentially the same results.  
}
From Fig. \ref{fig:jetAno}, we see that constrained re-calibration yields a model that captures the spreading rate of both plane and round jets.

\subsection{Airfoil Stall}
\label{sub:airfoil}

\begin{figure}
\centering
\includegraphics[width=0.4\textwidth]{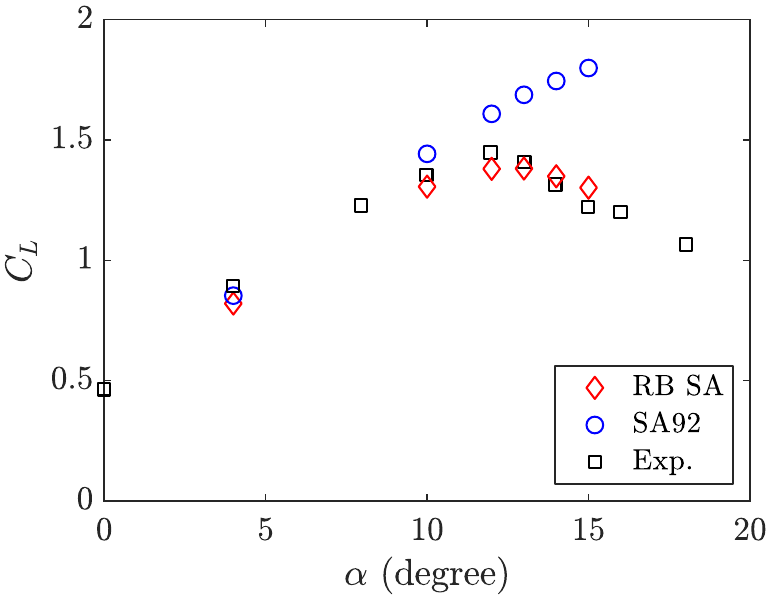}
\caption{Lift coefficient $C_L$ of a NACA4412 airfoil as a function of the angle of attack $\alpha$.
The experimental data in \cite{coles1979flying} is used as our reference.
The chord-length-based Reynolds number is at $Re_c=1.52\times10^6$.
}
\label{fig:naca}
\end{figure}

A stall manifests as a reduction in the lift coefficient $C_L = F_{\rm lift}/(0.5\rho U_{\rm inf}^2)$ as an air-foil increases its angle of attack \cite{ekaterinaris1998computational,sudharsan2022vorticity,sudharsan2023evaluating}.
Here, $F_{\rm lift}$ is the lift.
Predicting stall or the critical angle of attack is challenging. 
Figure \ref{fig:naca} shows the lift coefficient of a NACA4412 airfoil as a function of the angle of attack at $Re_c\equiv U_{\rm inf} c/\nu = 1.52\times10^6$, where $c$ is the chord length, and $U_{\rm inf}$ is the freestream velocity. 
We see that the baseline SA fails to capture stall at $\alpha\approx 12^\circ$.

Here, constrained re-calibration is attempted.
Applying BO, we obtain $c_{b1}=0.058$, $c_{s1}=0.985$, $c_{s2}=0.227$. 
The results of constrained re-calibration are shown in Fig. \ref{fig:naca}, and we see a close agreement between the RB SA model and the experimental data. 
Again, while any machine learning method, e.g., FIML, PIML, TBNN, can probably achieve this, the key is to achieve this without destroying the basic calibrations like the law of the wall.

\begin{figure}
\centering
\includegraphics[width=0.7\textwidth]{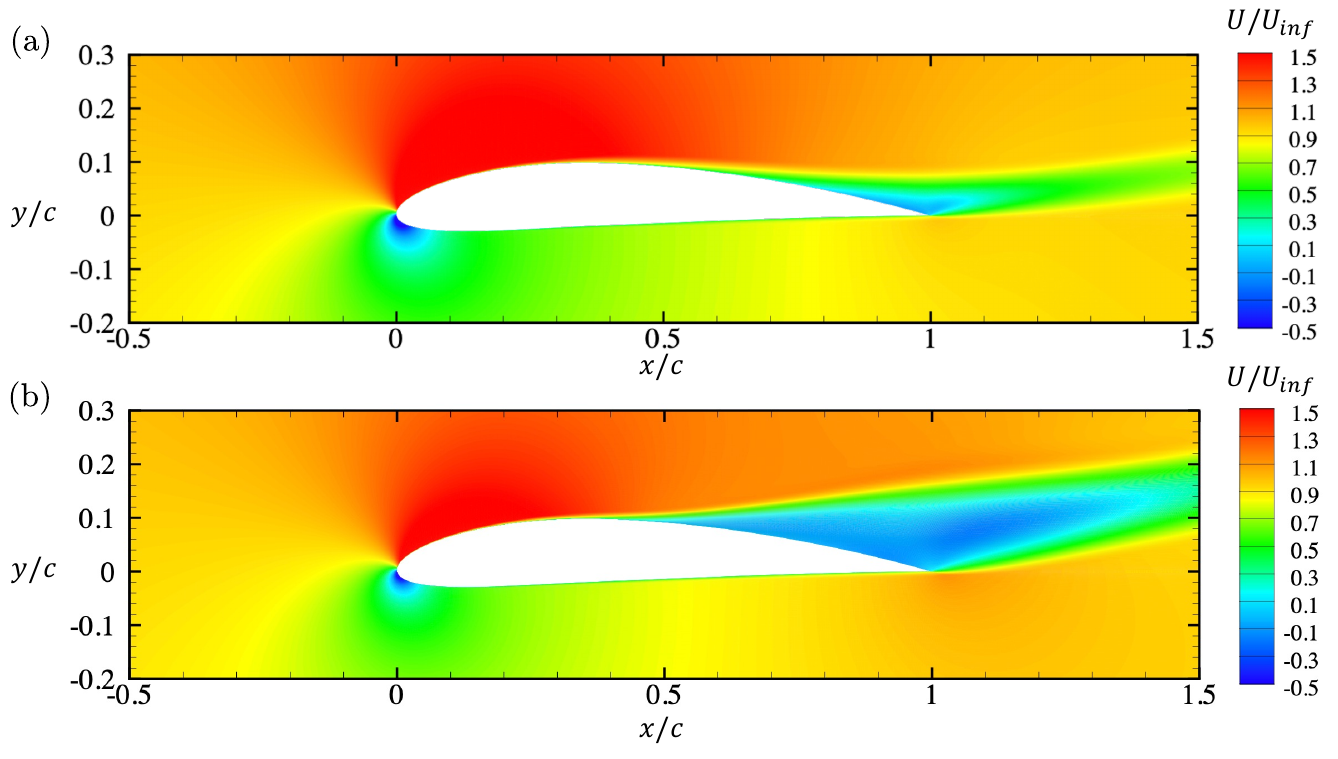}
\caption{Contours of $U$ at $\alpha=15^\circ$ predicted by (a) baseline SA (b) RB SA model.
}
\label{fig:N15}
\end{figure}

Furthermore, since we have developed a physical understanding of the adjustable constants, we can physically explain the results.
The difference in the baseline SA and the re-calibration can be attributed to their prediction of flow separation.
Figure \ref{fig:N15} shows the baseline SA and the RB SA predicted flow fields at $\alpha = 15^\circ$. 
We expect massive flow separation beyond the critical angle of attack, which is captured by the RB SA model but not the baseline SA model.

\subsection{Flow Recovery after separation}
\label{sub:recovery}

\begin{figure}
\centering
\includegraphics[width=0.8\textwidth]{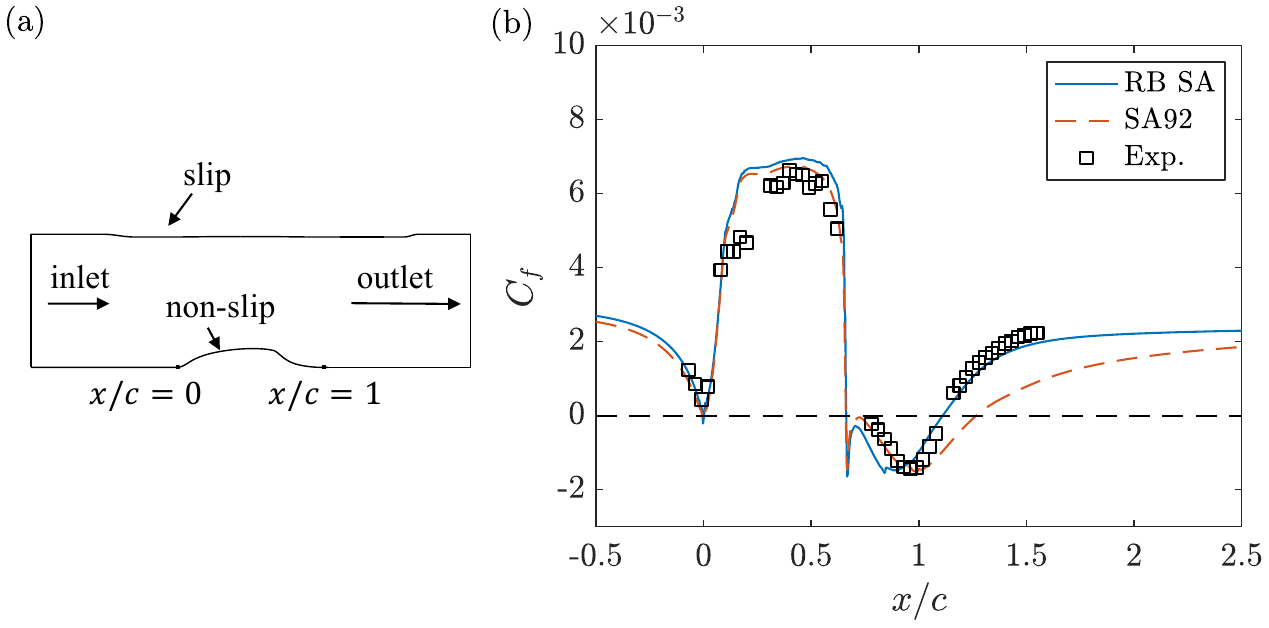}
\caption{(a) Schematic of the WMH case. 
(b) Skin friction coefficient $C_f$ as a function of the streamwise coordinate.
The experimental data in Ref. \cite{seifert2002active} is used as our reference.
The bump-length-based Reynolds number is $Re_{c}=9.36\times10^6$.
}
\label{fig:wmh}
\end{figure}

Flow recovery after separation is yet another long-standing issue in RANS modeling.
The baseline SA model is known to under-predict the recovery rate after separation.
Figure \ref{fig:wmh} (b) shows the skin friction coefficient in the wall-mounted hump (WMH) case.
A schematic of the flow is shown in figure \ref{fig:wmh} (a), and the reader is directed to the TMR site for further details of the case.
We see from figure \ref{fig:wmh} (b) that the baseline SA model under-predicts flow recovery and over-predicts the size of the separation bubble downstream of the hump at $Re_c=9.36\times10^6$. 
By training the model against the reference data,  we get $c_{b1}=0.217$, $c_{s1}=0.908$, $c_{s2}=1$,
and the constrained re-calibration result agrees closely with the experimental data.
Again, since the values of the adjustable constants are within their respective ranges, basic calibrations, including the fully-developed channel and flat-plate boundary layer, are not affected.  

\subsection{Secondary flows}
\label{sub:secondary}

\begin{figure}
\centering
\includegraphics[width=0.8\textwidth]{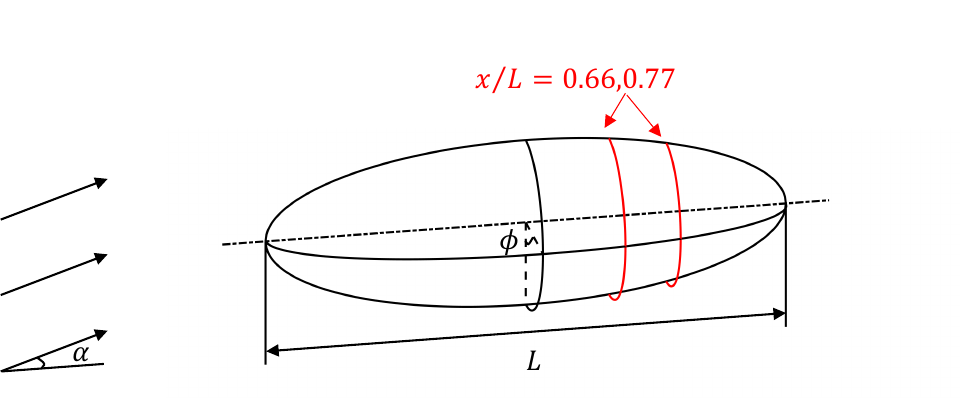}
\caption{Schematic of the 6:1 prolate spheroid at 20$^\circ$ angle of attack $\alpha$.
$\phi$ is the azimuthal angle.
}
\label{fig:psSketch}
\end{figure}

\begin{figure}
\centering
\includegraphics[width=0.8\textwidth]{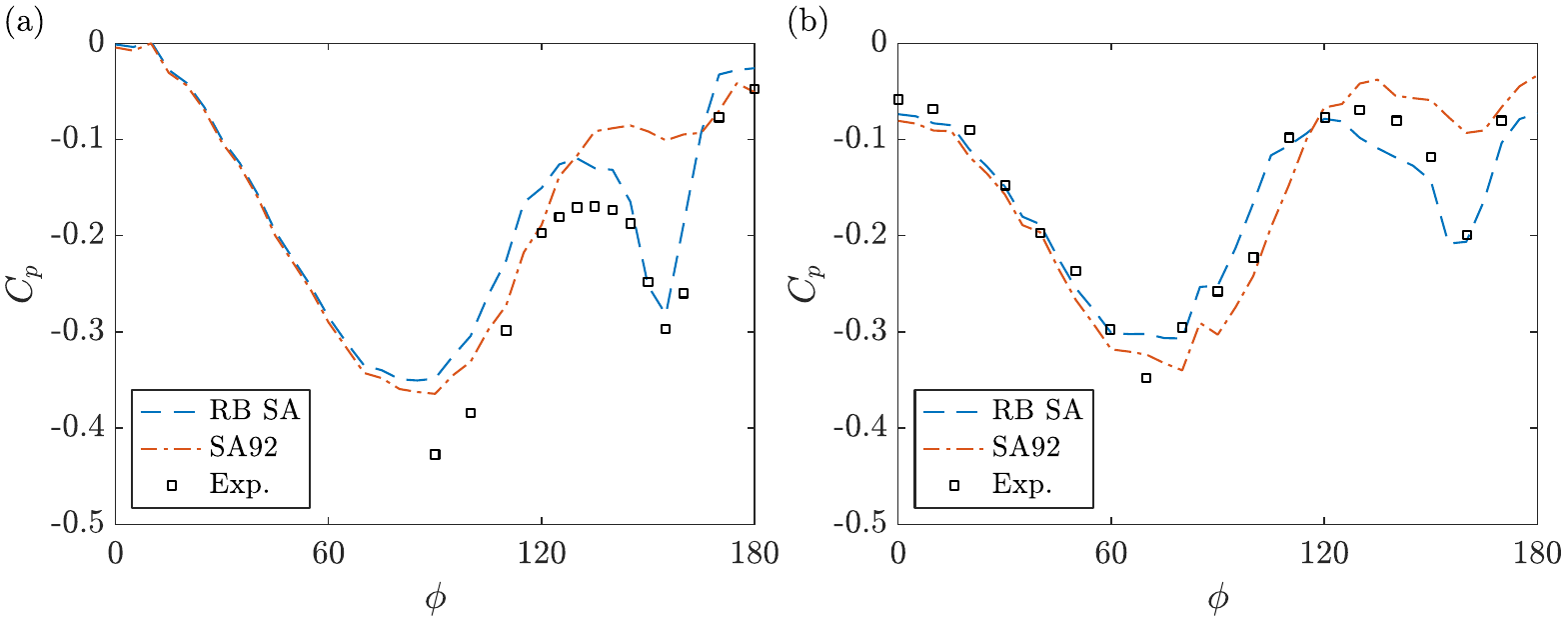}
\caption{Pressure coefficient $C_p$ as a function of the azimuthal angle $\phi$ at (a) $x/L=0.60$ (b) $x/L=0.77$.
}
\label{fig:psResults}
\end{figure}

In this subsection, we consider a 6:1 prolate spheroid at 20$^\circ$ angle of attack.
Figure \ref{fig:psSketch} shows a schematic of the flow.
The flow features a pair of counter-rotating vortices on the leeward side. 
The two vortices give rise to secondary flow separation on the surface of the prolate spheroid, which is hard to capture. 
Figure \ref{fig:psResults} shows the pressure coefficient $C_p \equiv (P-P_{\rm ref})/(0.5\rho U_{\rm inf}^2)$ as a function of the azimuthal angle at two locations, $x/L=0.66$ and $x/L=0.77$, as indicated by red color in figure \ref{fig:psSketch}, where $P$ is the absolute pressure, $P_{\rm ref}$ is the reference pressure, $\rho$ is the fluid density, and $U_{\rm inf}$ is the freestream velocity.
The experimental measurements in \cite{chesnakas1997detailed,wetzel1998measurement} are included as references.
The long-axis-based Reynolds number is $Re_L \equiv U_{\rm inf} L/\nu=4.2 \times 10^6$.
We see that the baseline SA model captures the primary separation but not the secondary separation.

\begin{figure}
\centering
\includegraphics[width=0.8\textwidth]{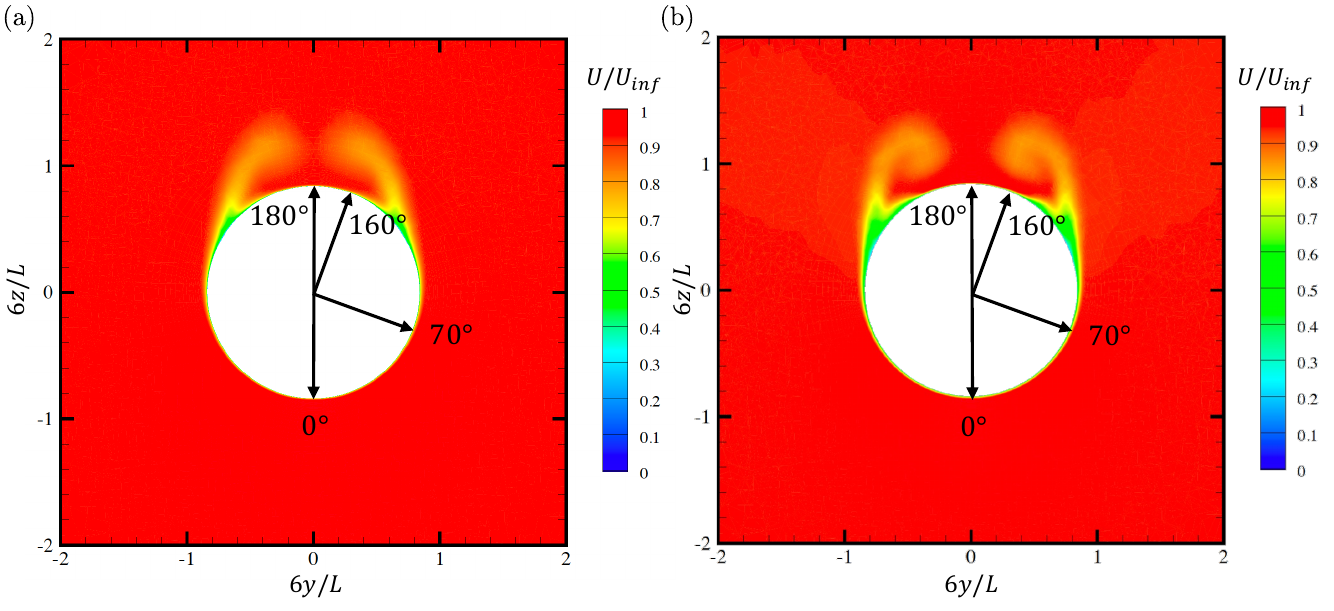}
\caption{Contour of the streamwise velocity at $x/L=0.77$. (a) baseline SA result, (b) RB SA model result.
}
\label{fig:psCon}
\end{figure}

By training the model against reference data, we have $c_{b1}=0.056$, $c_{s1}=0.311$, $c_{s2}=047$, and the RB SA model captures both the primary and the secondary separation. 
Again, since the values of the adjustable constants are well within their respective ranges, the channel and flat-plate calibrations are not affected.
The difference between the SA and the RB SA results can be attributed to their predictions of the secondary flow at the suction side.
Figure \ref{fig:psCon} shows the streamwise velocity at $x/L=0.77$.
We see that the RB SA model predicts more pronounced secondary flows, which in turn leads to the secondary separation on the surface.

\subsection{Robustness}
\label{sub:robust}

\begin{figure}
\centering
\includegraphics[width=\textwidth]{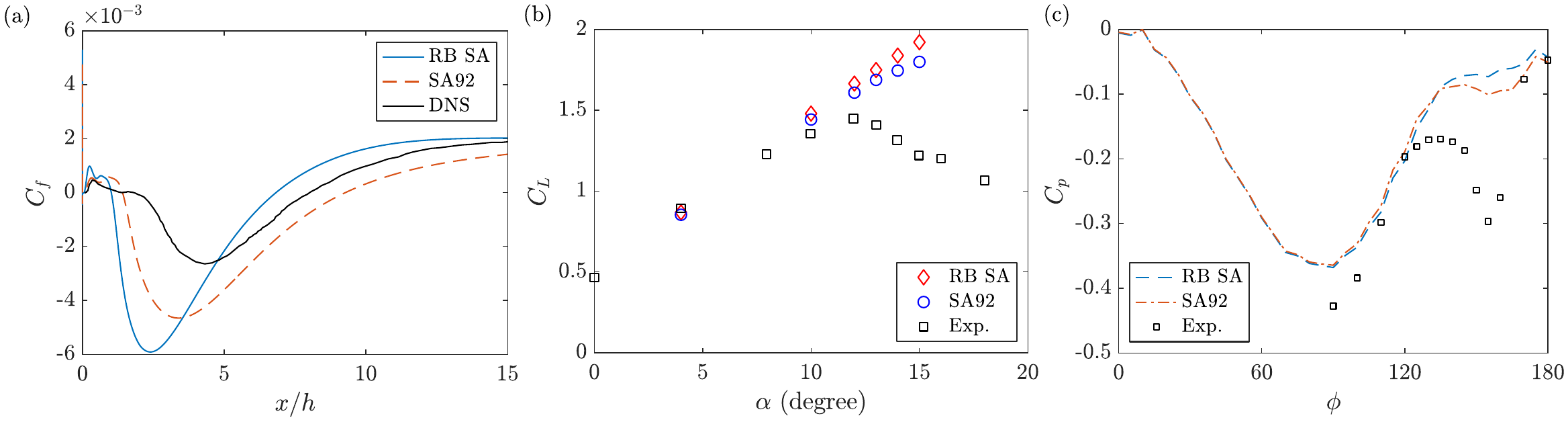}
\caption{Apply the constants calibrated for the WMH case to (a) the BFS case, (b) the NACA4412 case, and (c) the 6:1 prolate spheroid case.
(a) $C_f$ as a function  of $x$. 
(b) Lift coefficient $C_L$ as a function of the angle of attack $\alpha$ . 
(c) Pressure coefficient $C_p$ as a function of the azimuthal angle at $x/L=0.77$.}
\label{fig:allWithWMH}
\end{figure}

\begin{figure}
\centering
\includegraphics[width=0.4\textwidth]{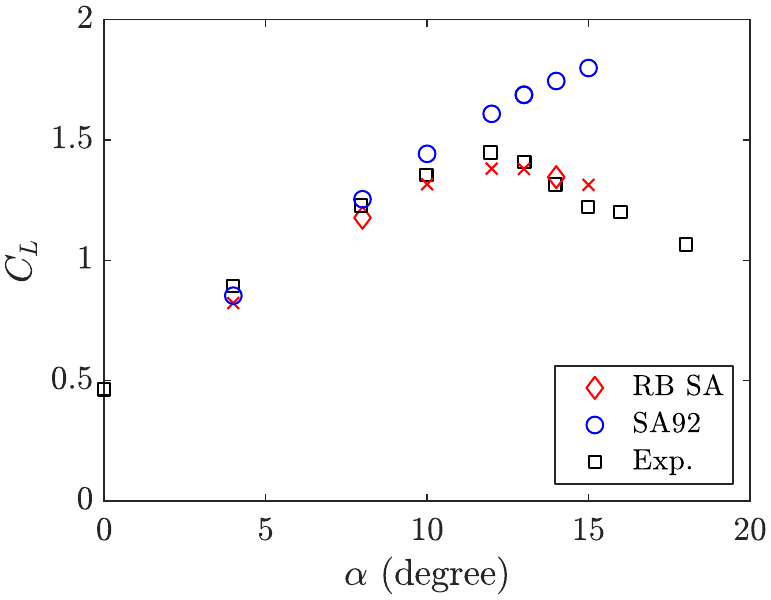}
\caption{Lift coefficient as a function of $\alpha$.
}
\label{fig:N447}
\end{figure}

The premise of the paper is that the generalizability of a RANS model comes from its basic calibrations.
In particular, we argue that as long as one preserves basic calibrations like the law of the wall, modifications to a baseline RANS model should not negatively impact its ability to generalize.
We have verified this conjecture in Sec. \ref{sec: validation}.
Here, we provide further evidence for improved robustness and generalizability.
Firstly, we take the values of the constants used for the WMH case and apply them to the BFS, NACA4412, and prolate spheroid cases.
The results are shown in figure \ref{fig:allWithWMH}.
We see that the constants that offer good performance for the WMH case do not destroy the baseline model outside the training data.
{\color{black}
Next, we repeat the exercise in Sec. \ref{sub:airfoil}, but instead of re-calibrating against data at all angles of attacks, here we re-calibrate against only two angles of attack: one pre-stall at $8^\circ$ and one post-stall at $14^\circ$.
The re-calibrated model is applied to airfoils at other angles of attack.
The hope is that by preserving the basic calibrations like the law of the wall, we will get models with good generalizability.
The results are shown in Figure \ref{fig:N447}.
We see that the re-calibrated model captures the lift coefficient at other angles of attack including the critical angle of attack.
}

\section{Conclusions}
\label{sec:conclusion}

Much of the existing work on the topic of data-enabled RANS modeling focuses on {\it how} one modifies a baseline RANS model, and not enough attention was given to {\it where} one should apply modifications.
This practice has caused robustness issues since the constants, functions, and terms in a RANS model are coupled, and changes to constants and functions in a RANS model often negatively impact its basic calibrations.
This paper focuses on {\it where} one should apply modifications.
We argue that modifications should only be introduced where they do not affect important calibrations of the baseline model.
To that end, we proposed constrained re-calibration, or the rubber-band approach. 
A rubber-band model contains degrees of freedom that do not affect designated calibrations of the baseline RANS model.
Adjusting these constants according to an existing machine-learning method, e.g., FIML or BO, allows one to accommodate specific needs while preserving the model's performance outside the training dataset.
We note that the method differs from including the designated calibrations in the training data, as the latter preserves specific flows, whereas the rubber-band approach preserves physics scalings.

To illustrate the approach, we re-formulated the Spalart-Allmaras model.
The model has four adjustable constants that control the model's behaviors in unbounded free shear flows, and non-equilibrium boundary-layer flows.
These adjustable constants are such that they do not affect the law of the wall.
We exploit the model and attempt to address some long-standing challenges in RANS modeling, including round-jet/plane-jet anomaly, airfoil stall, flow recovery after separation, and secondary flow separation.
The results are favorable.
Constrained re-calibration allows us to accommodate specific flows while maintaining the robustness of the baseline model outside the training dataset.

\section*{Acknowledgement}
Bin acknowledges NNSFC grant number 91752202.
Yang acknowledges ONR contract number N000142012315 and AFOSR award number FA9550-23-1-0272.

\bibliography{arxiv}

\begin{thebibliography}{69}
\newcommand{\enquote}[1]{``#1''}
\providecommand{\natexlab}[1]{#1}
\providecommand{\url}[1]{\texttt{#1}}
\providecommand{\urlprefix}{URL }
\expandafter\ifx\csname urlstyle\endcsname\relax
  \providecommand{\doi}[1]{\discretionary{}{}{}https://doi.org/#1}\else
  \providecommand{\doi}[1]{\discretionary{}{}{}\urlstyle{rm}\url{https://doi.org/#1}}\fi

\bibitem[{Durbin(2018)}]{durbin2018some}
Durbin, P.~A., \enquote{Some recent developments in turbulence closure
  modeling,} \emph{Ann. Rev. Fluid Mech.}, Vol.~50, 2018, pp. 77--103.
\newblock \doi{10.1146/annurev-fluid-122316-045020}.

\bibitem[{Pope and Pope(2000)}]{pope2000turbulent}
Pope, S.~B., and Pope, S.~B., \emph{Turbulent Flows}, Cambridge university
  press, 2000.

\bibitem[{Chapman(1979)}]{chapman1979computational}
Chapman, D.~R., \enquote{Computational aerodynamics development and outlook,}
  \emph{AIAA J.}, Vol.~17, No.~12, 1979, pp. 1293--1313.
\newblock \doi{10.2514/3.61311}.

\bibitem[{Yang and Griffin(2021)}]{yang2021grid}
Yang, X.~I., and Griffin, K.~P., \enquote{Grid-point and time-step requirements
  for direct numerical simulation and large-eddy simulation,} \emph{Physics of
  Fluids}, Vol.~33, No.~1, 2021, p. 015108.
\newblock \doi{10.1063/5.0036515}.

\bibitem[{Li et~al.(2022)Li, Yang, and Kunz}]{li2022grid}
Li, J.-Q.~J., Yang, X.~I., and Kunz, R.~F., \enquote{Grid-point and time-step
  requirements for large-eddy simulation and {R}eynolds-averaged
  {N}avier--{S}tokes of stratified wakes,} \emph{Phys. Fluids}, Vol.~34,
  No.~11, 2022, p. 115125.
\newblock \doi{10.1063/5.0127487}.

\bibitem[{Yang and Xia(2021)}]{yang2021bifurcation}
Yang, X.~I., and Xia, Z., \enquote{Bifurcation and multiple states in plane
  Couette flow with spanwise rotation,} \emph{Journal of Fluid Mechanics}, Vol.
  913, 2021, p. A49.

\bibitem[{Ravelet et~al.(2004)Ravelet, Mari{\'e}, Chiffaudel, and
  Daviaud}]{ravelet2004multistability}
Ravelet, F., Mari{\'e}, L., Chiffaudel, A., and Daviaud, F.,
  \enquote{Multistability and memory effect in a highly turbulent flow:
  Experimental evidence for a global bifurcation,} \emph{Physical review
  letters}, Vol.~93, No.~16, 2004, p. 164501.
\newblock \doi{10.1103/PhysRevLett.93.164501}.

\bibitem[{Xie et~al.(2018)Xie, Ding, and Xia}]{xie2018flow}
Xie, Y.-C., Ding, G.-Y., and Xia, K.-Q., \enquote{Flow topology transition via
  global bifurcation in thermally driven turbulence,} \emph{Physical Review
  Letters}, Vol. 120, No.~21, 2018, p. 214501.
\newblock \doi{10.1103/PhysRevLett.120.214501}.

\bibitem[{Weiss et~al.(2010)Weiss, Stevens, Zhong, Clercx, Lohse, and
  Ahlers}]{weiss2010finite}
Weiss, S., Stevens, R.~J., Zhong, J.-Q., Clercx, H.~J., Lohse, D., and Ahlers,
  G., \enquote{Finite-size effects lead to supercritical bifurcations in
  turbulent rotating Rayleigh-B{\'e}nard convection,} \emph{Physical review
  letters}, Vol. 105, No.~22, 2010, p. 224501.
\newblock \doi{10.1103/PhysRevLett.105.224501}.

\bibitem[{Spalart(2015)}]{spalart2015philosophies}
Spalart, P.~R., \enquote{Philosophies and fallacies in turbulence modeling,}
  \emph{Prog. Aerosp. Sci.}, Vol.~74, 2015, pp. 1--15.
\newblock \doi{10.1016/j.paerosci.2014.12.004}.

\bibitem[{Spalart and Allmaras(1992)}]{spalart1992one}
Spalart, P., and Allmaras, S., \enquote{A one-equation turbulence model for
  aerodynamic flows,} \emph{30th Aerospace Sciences Meeting and Exhibit}, 1992,
  p. 439.
\newblock \doi{10.2514/6.1992-439}.

\bibitem[{Menter(1994)}]{menter1994two}
Menter, F.~R., \enquote{Two-equation eddy-viscosity turbulence models for
  engineering applications,} \emph{AIAA J.}, Vol.~32, No.~8, 1994, pp.
  1598--1605.
\newblock \doi{10.2514/3.12149}.

\bibitem[{Wilcox(1988)}]{wilcox1988reassessment}
Wilcox, D.~C., \enquote{Reassessment of the scale-determining equation for
  advanced turbulence models,} \emph{AIAA J.}, Vol.~26, No.~11, 1988, pp.
  1299--1310.
\newblock \doi{10.2514/3.10041}.

\bibitem[{Wilcox et~al.(1998)}]{wilcox1998turbulence}
Wilcox, D.~C., et~al., \emph{Turbulence Modeling for CFD}, Vol.~2, DCW
  industries La Canada, CA, 1998.

\bibitem[{Wilcox(2008)}]{wilcox2008formulation}
Wilcox, D.~C., \enquote{Formulation of the $k$-$\omega$ turbulence model
  revisited,} \emph{AIAA J.}, Vol.~46, No.~11, 2008, pp. 2823--2838.
\newblock \doi{10.2514/1.36541}.

\bibitem[{Chien(1982)}]{chien1982predictions}
Chien, K.-Y., \enquote{Predictions of channel and boundary-layer flows with a
  low-Reynolds-number turbulence model,} \emph{AIAA J.}, Vol.~20, No.~1, 1982,
  pp. 33--38.
\newblock \doi{doi.org/10.2514/3.51043}.

\bibitem[{Launder and Spalding(1983)}]{launder1983numerical}
Launder, B.~E., and Spalding, D.~B., \enquote{The Numerical Computation of
  Turbulent Flows,} \emph{Numerical Prediction of Flow, Heat transfer,
  Turbulence and Combustion}, Elsevier, 1983, pp. 96--116.
\newblock \doi{10.1016/B978-0-08-030937-8.50016-7}.

\bibitem[{Abdol-Hamid et~al.(2016)Abdol-Hamid, Carlson, and
  Rumsey}]{abdol2016verification}
Abdol-Hamid, K.~S., Carlson, J.-R., and Rumsey, C.~L., \enquote{Verification
  and Validation of the k-kL Turbulence Model in FUN3D and CFL3D Codes,}
  \emph{46th AIAA Fluid Dynamics Conference}, 2016, p. 3941.
\newblock \doi{10.2514/6.2016-3941}.

\bibitem[{Durbin(1991)}]{durbin1991near}
Durbin, P.~A., \enquote{Near-wall turbulence closure modeling without
  “damping functions”,} \emph{Theor Comp Fluid Dyn}, Vol.~3, No.~1, 1991,
  pp. 1--13.
\newblock \doi{10.1007/BF00271513}.

\bibitem[{Huang et~al.(2023)Huang, Chyczewski, Xia, Kunz, and
  Yang}]{huang2023distilling}
Huang, X., Chyczewski, T., Xia, Z., Kunz, R., and Yang, X., \enquote{Distilling
  experience into a physically interpretable recommender system for
  computational model selection,} \emph{Sci. Rep.}, Vol.~13, No.~1, 2023, p.
  2225.
\newblock \doi{10.1038/s41598-023-27426-5}.

\bibitem[{Zhang et~al.(2020)Zhang, He, Xie, Xiao, and
  Tian}]{zhang2020methodology}
Zhang, Y.-s., He, Z.-w., Xie, H.-s., Xiao, M.-J., and Tian, B.-l.,
  \enquote{Methodology for determining coefficients of turbulent mixing model,}
  \emph{J. Fluid Mech.}, Vol. 905, 2020, p. A26.
\newblock \doi{10.1017/jfm.2020.726}.

\bibitem[{Parente et~al.(2011)Parente, Gorl{\'e}, Van~Beeck, and
  Benocci}]{parente2011improved}
Parente, A., Gorl{\'e}, C., Van~Beeck, J., and Benocci, C., \enquote{Improved
  k--$\varepsilon$ model and wall function formulation for the RANS simulation
  of ABL flows,} \emph{J. Wind Eng Ind Aerod}, Vol.~99, No.~4, 2011, pp.
  267--278.
\newblock \doi{10.1016/j.jweia.2010.12.017}.

\bibitem[{Cindori et~al.(2018)Cindori, Jureti{\'c}, Kozmar, and
  D{\v{z}}ijan}]{cindori2018steady}
Cindori, M., Jureti{\'c}, F., Kozmar, H., and D{\v{z}}ijan, I., \enquote{Steady
  RANS model of the homogeneous atmospheric boundary layer,} \emph{J. Wind Eng
  Ind Aerod}, Vol. 173, 2018, pp. 289--301.
\newblock \doi{10.1016/j.jweia.2017.12.006}.

\bibitem[{Gimenez and Bre(2019)}]{gimenez2019optimization}
Gimenez, J.~M., and Bre, F., \enquote{Optimization of RANS turbulence models
  using genetic algorithms to improve the prediction of wind pressure
  coefficients on low-rise buildings,} \emph{J. Wind Eng Ind Aerod}, Vol. 193,
  2019, p. 103978.
\newblock \doi{10.1016/j.jweia.2019.103978}.

\bibitem[{Ling et~al.(2016{\natexlab{a}})Ling, Jones, and
  Templeton}]{ling2016machine}
Ling, J., Jones, R., and Templeton, J., \enquote{Machine learning strategies
  for systems with invariance properties,} \emph{J Comput Phys}, Vol. 318,
  2016{\natexlab{a}}, pp. 22--35.
\newblock \doi{10.1016/j.jcp.2016.05.003}.

\bibitem[{Ling et~al.(2016{\natexlab{b}})Ling, Kurzawski, and
  Templeton}]{ling2016reynolds}
Ling, J., Kurzawski, A., and Templeton, J., \enquote{Reynolds averaged
  turbulence modelling using deep neural networks with embedded invariance,}
  \emph{J. Fluid Mech.}, Vol. 807, 2016{\natexlab{b}}, pp. 155--166.
\newblock \doi{10.1017/jfm.2016.615}.

\bibitem[{Wang et~al.(2017)Wang, Wu, and Xiao}]{wang2017physics}
Wang, J.-X., Wu, J.-L., and Xiao, H., \enquote{Physics-informed machine
  learning approach for reconstructing {R}eynolds stress modeling discrepancies
  based on {DNS} data,} \emph{Phys. Rev. Fluids}, Vol.~2, No.~3, 2017, p.
  034603.
\newblock \doi{10.1103/PhysRevFluids.2.034603}.

\bibitem[{Wu et~al.(2018)Wu, Xiao, and Paterson}]{wu2018physics}
Wu, J.-L., Xiao, H., and Paterson, E., \enquote{Physics-informed machine
  learning approach for augmenting turbulence models: A comprehensive
  framework,} \emph{Phys. Rev. Fluids}, Vol.~3, No.~7, 2018, p. 074602.
\newblock \doi{10.1103/PhysRevFluids.3.074602}.

\bibitem[{Singh and Duraisamy(2016)}]{singh2016using}
Singh, A.~P., and Duraisamy, K., \enquote{Using field inversion to quantify
  functional errors in turbulence closures,} \emph{Phys. Fluids}, Vol.~28,
  No.~4, 2016, p. 045110.
\newblock \doi{10.1063/1.4947045}.

\bibitem[{Singh et~al.(2017)Singh, Medida, and Duraisamy}]{singh2017machine}
Singh, A.~P., Medida, S., and Duraisamy, K.,
  \enquote{Machine-learning-augmented predictive modeling of turbulent
  separated flows over airfoils,} \emph{AIAA J.}, Vol.~55, No.~7, 2017, pp.
  2215--2227.
\newblock \doi{10.2514/1.J055595}.

\bibitem[{Fang et~al.(2023)Fang, Zhao, Waschkowski, Ooi, and
  Sandberg}]{fang2023toward}
Fang, Y., Zhao, Y., Waschkowski, F., Ooi, A.~S., and Sandberg, R.~D.,
  \enquote{Toward More General Turbulence Models via Multicase
  Computational-Fluid-Dynamics-Driven Training,} \emph{AIAA J.}, 2023, pp.
  1--16.
\newblock \doi{10.2514/1.J062572}.

\bibitem[{Zhao et~al.(2020)Zhao, Akolekar, Weatheritt, Michelassi, and
  Sandberg}]{zhao2020rans}
Zhao, Y., Akolekar, H.~D., Weatheritt, J., Michelassi, V., and Sandberg, R.~D.,
  \enquote{{RANS} turbulence model development using {CFD}-driven machine
  learning,} \emph{J Comput Phys}, Vol. 411, 2020, p. 109413.
\newblock \doi{10.1016/j.jcp.2020.109413}.

\bibitem[{Duraisamy et~al.(2019)Duraisamy, Iaccarino, and
  Xiao}]{duraisamy2019turbulence}
Duraisamy, K., Iaccarino, G., and Xiao, H., \enquote{Turbulence modeling in the
  age of data,} \emph{Ann. Rev. Fluid Mech.}, Vol.~51, 2019, pp. 357--377.
\newblock \doi{10.1146/annurev-fluid-010518-040547}.

\bibitem[{Rumsey et~al.(2022)Rumsey, Coleman, and Wang}]{rumsey2022search}
Rumsey, C.~L., Coleman, G.~N., and Wang, L., \enquote{In search of data-driven
  improvements to {RANS} models applied to separated flows,} \emph{AIAA SCITECH
  2022 Forum}, 2022, p. 0937.
\newblock \doi{10.2514/6.2022-0937}.

\bibitem[{Ferrero et~al.(2020)Ferrero, Iollo, and Larocca}]{ferrero2020field}
Ferrero, A., Iollo, A., and Larocca, F., \enquote{Field inversion for
  data-augmented {RANS} modelling in turbomachinery flows,} \emph{Computers \&
  Fluids}, Vol. 201, 2020, p. 104474.
\newblock \doi{10.1016/j.compfluid.2020.104474}.

\bibitem[{Wu and Zhang(2023)}]{wu2023enhancing}
Wu, C., and Zhang, Y., \enquote{Enhancing the {SST} Turbulence Model with
  Symbolic Regression: A Generalizable and Interpretable Data-Driven Approach,}
  \emph{arXiv preprint arXiv:2304.11347}, 2023.
\newblock \doi{10.48550/arXiv.2304.11347}.

\bibitem[{Spalart(2023)}]{spalart2023old}
Spalart, P., \enquote{An Old-Fashioned Framework for Machine Learning in
  Turbulence Modeling,} \emph{arXiv preprint arXiv:2308.00837}, 2023.

\bibitem[{Menter et~al.(2019)Menter, Lechner, and Matyushenko}]{menter2019best}
Menter, F., Lechner, R., and Matyushenko, A., \enquote{Best practice:
  generalized k-$\omega$ two-equation turbulence model in ANSYS CFD (GEKO),}
  \emph{ANSYS Germany GmbH}, 2019.

\bibitem[{Strokach et~al.(2021)Strokach, Zhukov, Borovik, Sternin, and
  Haidn}]{strokach2021simulation}
Strokach, E., Zhukov, V., Borovik, I., Sternin, A., and Haidn, O.~J.,
  \enquote{Simulation of a gox-gch4 rocket combustor and the effect of the geko
  turbulence model coefficients,} \emph{Aerospace}, Vol.~8, No.~11, 2021, p.
  341.
\newblock \doi{10.3390/aerospace8110341}.

\bibitem[{Szudarek et~al.(2022)Szudarek, Piechna, Prusi{\'n}ski, and
  Rudniak}]{szudarek2022cfd}
Szudarek, M., Piechna, A., Prusi{\'n}ski, P., and Rudniak, L., \enquote{CFD
  Study of high-speed train in crosswinds for large yaw angles with RANS-based
  turbulence models including GEKO tuning approach,} \emph{Energies}, Vol.~15,
  No.~18, 2022, p. 6549.
\newblock \doi{10.3390/en15186549}.

\bibitem[{Jung et~al.(2021)Jung, Chang, and Bae}]{jung2021uncertainty}
Jung, Y.-K., Chang, K., and Bae, J.~H., \enquote{Uncertainty Quantification of
  GEKO Model Coefficients on Compressible Flows,} \emph{Int. J. Aerosp. Eng.},
  Vol. 2021, 2021, pp. 1--17.
\newblock \doi{10.1155/2021/9998449}.

\bibitem[{Sharkey and Menter(2019)}]{sharkey2019numerical}
Sharkey, P., and Menter, F., \enquote{A numerical investigation of the
  turbulent flow around a scale model JBC hull using the Generalized k-omega
  ({GEKO}) turbulence model,} \emph{11th International Workshop on Ship and
  Marine Hydrodynamics (IWSH2019)}, 2019.

\bibitem[{Nair and Mathew(2022)}]{nair2022resistance}
Nair, A.~S., and Mathew, M., \enquote{Resistance estimation of ships using GEKO
  turbulence model in ANSYS Fluent,} \emph{OCEANS 2022-Chennai}, IEEE, 2022,
  pp. 1--9.

\bibitem[{Bridges and Wernet(2010)}]{bridges2010establishing}
Bridges, J., and Wernet, M., \enquote{Establishing consensus turbulence
  statistics for hot subsonic jets,} \emph{16th AIAA/CEAS aeroacoustics
  conference}, 2010, p. 3751.
\newblock \doi{10.2514/6.2010-3751}.

\bibitem[{Bridges and Wernet(2011)}]{bridges2011nasa}
Bridges, J., and Wernet, M.~P., \enquote{The NASA subsonic jet particle image
  velocimetry ({PIV}) dataset,} , 2011.

\bibitem[{Barri et~al.(2010)Barri, El~Khoury, Andersson, and
  Pettersen}]{barri2010dns}
Barri, M., El~Khoury, G.~K., Andersson, H.~I., and Pettersen, B.,
  \enquote{{DNS} of backward-facing step flow with fully turbulent inflow,}
  \emph{Int J Numer Methods Fluids}, Vol.~64, No.~7, 2010, pp. 777--792.
\newblock \doi{10.1002/fld.2176}.

\bibitem[{Seifert and Pack(2002)}]{seifert2002active}
Seifert, A., and Pack, L.~G., \enquote{Active flow separation control on
  wall-mounted hump at high {R}eynolds numbers,} \emph{AIAA J.}, Vol.~40,
  No.~7, 2002, pp. 1363--1372.
\newblock \doi{10.2514/2.1796}.

\bibitem[{Xiao et~al.(2007)Xiao, Zhang, Huang, Chen, and
  Fu}]{xiao2007prediction}
Xiao, Z., Zhang, Y., Huang, J., Chen, H., and Fu, S., \enquote{Prediction of
  separation flows around a 6: 1 prolate spheroid using RANS/LES hybrid
  approaches,} \emph{Am. Meteorol. Soc.}, Vol.~23, No.~4, 2007, pp. 369--382.
\newblock \doi{10.1007/s10409-007-0073-6}.

\bibitem[{Vadrot et~al.(2023)Vadrot, Yang, Bae, and Abkar}]{vadrot2023log}
Vadrot, A., Yang, X.~I., Bae, H.~J., and Abkar, M., \enquote{Log-law recovery
  through reinforcement-learning wall model for large eddy simulation,}
  \emph{Phys. Fluids}, Vol.~35, No.~5, 2023.
\newblock \doi{10.1063/5.0147570}.

\bibitem[{Bin et~al.(2023)Bin, George, and Yang}]{bin2023data}
Bin, Y., George, and Yang, X. I.~A., \enquote{A data-enabled re-calibration of
  the Spalart-Allmaras model for general purposes,} \emph{AIAA J.}, 2023.

\bibitem[{Mellor and Herring(1968)}]{mellor1968two}
Mellor, G.~L., and Herring, H., \enquote{Two methods of calculating turbulent
  boundary layer behavior based on numerical solutions of the equations of
  motion,} \emph{Proc. Conf. Turb. Boundary Layer Pred., Stanford}, 1968.

\bibitem[{Baldwin and Barth(1990)}]{baldwin1990one}
Baldwin, B., and Barth, T., \enquote{A one-equation turbulence transport model
  for high Reynolds number wall-bounded flows,} \emph{NASA TM 102847}, 1990.

\bibitem[{Baldwin and Barth(1991)}]{baldwin1991one}
Baldwin, B., and Barth, T., \enquote{A one-equation turbulence transport model
  for high Reynolds number wall-bounded flows,} \emph{29th aerospace sciences
  meeting}, 1991, p. 610.
\newblock \doi{10.2514/6.1991-610}.

\bibitem[{Lee and Moser(2015)}]{lee2015direct}
Lee, M., and Moser, R.~D., \enquote{Direct numerical simulation of turbulent
  channel flow up to ${R}e_\tau=5200$,} \emph{J. Fluid Mech.}, Vol. 774, 2015,
  pp. 395--415.
\newblock \doi{10.1017/jfm.2015.268}.

\bibitem[{Witze and Dwyer(1976)}]{witze1976turbulent}
Witze, P., and Dwyer, H., \enquote{The turbulent radial jet,} \emph{J. Fluid
  Mech.}, Vol.~75, No.~3, 1976, pp. 401--417.
\newblock \doi{10.1017/S0022112076000293}.

\bibitem[{Bradbury(1965)}]{bradbury1965structure}
Bradbury, L., \enquote{The structure of a self-preserving turbulent plane jet,}
  \emph{J. Fluid Mech.}, Vol.~23, No.~1, 1965, pp. 31--64.
\newblock \doi{10.1017/S0022112065001222}.

\bibitem[{Coles and Wadcock(1979)}]{coles1979flying}
Coles, D., and Wadcock, A.~J., \enquote{Flying-hot-wire study of flow past an
  NACA 4412 airfoil at maximum lift,} \emph{AIAA J.}, Vol.~17, No.~4, 1979, pp.
  321--329.
\newblock \doi{10.2514/3.61127}.

\bibitem[{Chesnakas and Simpson(1997)}]{chesnakas1997detailed}
Chesnakas, C.~J., and Simpson, R.~L., \enquote{Detailed investigation of the
  three-dimensional separation about a 6: 1 prolate spheroid,} \emph{AIAA J.},
  Vol.~35, No.~6, 1997, pp. 990--999.
\newblock \doi{10.2514/2.208}.

\bibitem[{Wetzel et~al.(1998)Wetzel, Simpson, and
  Chesnakas}]{wetzel1998measurement}
Wetzel, T.~G., Simpson, R.~L., and Chesnakas, C.~J., \enquote{Measurement of
  three-dimensional crossflow separation,} \emph{AIAA J.}, Vol.~36, No.~4,
  1998, pp. 557--564.
\newblock \doi{10.2514/2.429}.

\bibitem[{Jasak et~al.(2007)Jasak, Jemcov, Tukovic et~al.}]{jasak2007openfoam}
Jasak, H., Jemcov, A., Tukovic, Z., et~al., \enquote{OpenFOAM: A C++ library
  for complex physics simulations,} \emph{International Workshop on Coupled
  Methods in Numerical Dynamics}, Vol. 1000, 2007, pp. 1--20.

\bibitem[{Git(2023)}]{GitHubLink}
\enquote{{R}ubber {B}and {S}palart-{A}llmaras model,} , 2023.
\newblock
  \urlprefix\url{https://github.com/yuanweibin/Rubber-Band-Spalart-Allmaras}.

\bibitem[{Amiri et~al.(2019)Amiri, Vitola, Sphaier, and
  Esperan{\c{c}}a}]{amiri2019rans}
Amiri, M.~M., Vitola, M.~A., Sphaier, S.~H., and Esperan{\c{c}}a, P.~T.,
  \enquote{RANS feasibility study of using roughness to mimic transition strip
  effect on the crossflowseparation over a 6: 1 prolate-spheroid,} \emph{J
  Hydrodynam B}, Vol.~31, 2019, pp. 570--581.

\bibitem[{Pope(1978)}]{pope1978explanation}
Pope, S., \enquote{An explanation of the turbulent round-jet/plane-jet
  anomaly,} \emph{AIAA J.}, Vol.~16, No.~3, 1978, pp. 279--281.
\newblock \doi{10.2514/3.7521}.

\bibitem[{Bardina et~al.(1997)Bardina, Huang, and
  Coakley}]{bardina1997turbulence}
Bardina, J.~E., Huang, P.~G., and Coakley, T.~J., \enquote{Turbulence Modeling
  Validation, Testing, and Development,} Tech. rep., 1997.

\bibitem[{Wygnanski and Fiedler(1969)}]{wygnanski1969some}
Wygnanski, I., and Fiedler, H., \enquote{Some measurements in the
  self-preserving jet,} \emph{J. Fluid Mech.}, Vol.~38, No.~3, 1969, pp.
  577--612.
\newblock \doi{10.1017/S0022112069000358}.

\bibitem[{Hussein et~al.(1994)Hussein, Capp, and George}]{hussein1994velocity}
Hussein, H.~J., Capp, S.~P., and George, W.~K., \enquote{Velocity measurements
  in a high-{R}eynolds-number, momentum-conserving, axisymmetric, turbulent
  jet,} \emph{Journal of Fluid Mechanics}, Vol. 258, 1994, pp. 31--75.
\newblock \doi{10.1017/S002211209400323X}.

\bibitem[{Ekaterinaris and Platzer(1998)}]{ekaterinaris1998computational}
Ekaterinaris, J.~A., and Platzer, M.~F., \enquote{Computational prediction of
  airfoil dynamic stall,} \emph{Prog. Aerosp. Sci.}, Vol.~33, No. 11-12, 1998,
  pp. 759--846.
\newblock \doi{10.1016/S0376-0421(97)00012-2}.

\bibitem[{Sudharsan et~al.(2022)Sudharsan, Ganapathysubramanian, and
  Sharma}]{sudharsan2022vorticity}
Sudharsan, S., Ganapathysubramanian, B., and Sharma, A., \enquote{A
  vorticity-based criterion to characterise leading edge dynamic stall onset,}
  \emph{J. Fluid Mech.}, Vol. 935, 2022, p. A10.
\newblock \doi{10.1017/jfm.2021.1149}.

\bibitem[{Sudharsan et~al.(2023)Sudharsan, Narsipur, and
  Sharma}]{sudharsan2023evaluating}
Sudharsan, S., Narsipur, S., and Sharma, A., \enquote{Evaluating Dynamic
  Stall-Onset Criteria for Mixed and Trailing-Edge Stall,} \emph{AIAA J.},
  Vol.~61, No.~3, 2023, pp. 1181--1196.
\newblock \doi{10.2514/1.J062011}.

\end{thebibliography}

\end{document}